\documentclass[aps,prl,twocolumn,superscriptaddress,nobibnotes,amsmath,amssymb]{revtex4-2}
\usepackage{graphicx}
\usepackage{dcolumn}
\usepackage{bm}
\usepackage{natbib}
\usepackage{color}
\usepackage{amsmath}
\usepackage[]{hyperref}
\hypersetup{
    colorlinks=true,
    allcolors = {blue},
    }
\usepackage[normalem]{ulem}

\begin{document}

\title{Fate of the superconducting state in floating islands of hybrid nanowire devices}
\author{E.V.~Shpagina}\affiliation{Osipyan Institute of Solid State Physics, Russian Academy of
Sciences, 142432 Chernogolovka, Russian Federation}
\affiliation{National Research University Higher School of Economics, 20 Myasnitskaya Street, 101000 Moscow, Russian Federation}
\author{E.S.~Tikhonov}\affiliation{Osipyan Institute of Solid State Physics, Russian Academy of
Sciences, 142432 Chernogolovka, Russian Federation}
\affiliation{National Research University Higher School of Economics, 20 Myasnitskaya Street, 101000 Moscow, Russian Federation}
\author{D.~Ruhstorfer}
\affiliation{Walter Schottky Institut, Physik Department, and Center for Nanotechnology and Nanomaterials, Technische Universit\"{a}t M\"{u}nchen, Am Coulombwall 4, Garching 85748, Germany}
\author{G.~Koblm\"{u}ller}
\affiliation{Walter Schottky Institut, Physik Department, and Center for Nanotechnology and Nanomaterials, Technische Universit\"{a}t M\"{u}nchen, Am Coulombwall 4, Garching 85748, Germany}
\author{V.S.~Khrapai}
\affiliation{Osipyan Institute of Solid State Physics, Russian Academy of
Sciences, 142432 Chernogolovka, Russian Federation}
\affiliation{National Research University Higher School of Economics, 20 Myasnitskaya Street, 101000 Moscow, Russian Federation}
\email{dick@issp.ac.ru}

\begin{abstract} We investigate the impact of transport current on the superconducting order parameter in superconducting islands in full-shell epitaxial Al-InAs nanowires. Depending on a device layout, the suppression of superconductivity occurs in three fundamentally different ways -- by a critical current in the case of superconducting reservoirs and by a critical voltage or by a critical Joule power in the case of normal reservoirs. In the latter case, the collapse of the superconducting state depends on the ratio of the dwell time and the electron-phonon relaxation time of quasiparticles in the island. For low resistive and high resistive coupling to the reservoirs, respectively, the relaxation-free regime and the strong electron-phonon relaxation regime are realized. Our results shed light on potential shortcomings of finite-bias transport spectroscopy in floating islands. 
\end{abstract}

\maketitle


Hybrid semiconductor-superconductor (semi-super) nanowires (NWs) are a lively research topic on the
superconducting proximity effect, especially in its modern forefront --- the Majorana research~\cite{lutchyn2010,oreg2010,prada2020}. 
Such devices are investigated with the main focus on the semiconductor side and the
quantities of interest include the induced spectral gap in the NW~\cite{chang2015,krogstrup2015,gul2017,bubis2017,junger2020,yu2023} and Andreev bound states energies~\cite{junger2019,junger2020}, the non-local response~\cite{stanescu2014,rosdahl2018,lai2019,menard2020,denisov2021,puglia2021,pan2021,denisov2022,wang2022a,kejriwal2022} and sub-gap heat conductance~\cite{akhmerov2011,denisov2021,pan2021,denisov2022}, the Josephson effect~\cite{nishio2011,abay2012,abay2014,paajaste2015,perla2021,kousar2022,spanton2017,hart2019}, the variety of zero bias conductance anomalies~\cite{das2012,mourik2012,vaitiekenas2020a,vaitiekenas2021,valentini2021} and Cooper-pair splitting~\cite{hofstetter2009,herrmann2010,das2012a,baba2018,wang2022,bordoloi2022,scherubl2022}. Since typical currents in semi-super hybrids are orders of magnitude smaller than the critical current of the superconductor, the order parameter ($\Delta$) is rarely a target for the experimentalists beyond the equilibrium characterization~\cite{vaitiekenas2020}.  As we show here, quasiparticle non-equilibrium and relaxation are much more relevant than current for the superconductivity in such devices.

In semi-super hybrids with a mesoscopic superconductor, which is not a part of the superconducting reservoir, referred to as the floating S-island below, the non-equilibrium mediated by the finite bias voltage ($V$) leads to a twofold complication. First, the quasiparticle population interplays with $\Delta$, since they are bound in the Bardeen-Cooper-Schrieffer (BCS) theory~\cite{keizer2006,snyman2009}. Second, this interplay may itself depend on the inelastic relaxation, provided quasiparticles spend enough time in the island~\cite{huard2007,sivre2018,rosenblatt2020}. Known in all-metal devices~\cite{vercruyssen2012}, non-equilibrium effects are not discussed in semi-super hybrids~\cite{fu2010,ulrich2015,albrecht2016,lai2021,hao2022,souto2022,valentini2022}, with rare exceptions~\cite{roddaro2011,bubis2021}. Clear indications of non-equilibrium effects were recently found in hybrid NWs at high biases $|V|\gg\Delta/e$~\cite{liu2023, ibabe2023}. The microscopic role of the energy relaxation in these experiments, however, remains hidden. 
%
%
%

In this article we investigate the interplay of quasiparticle non-equilibrium, superconductivity and electron-phonon (e-ph) relaxation in epitaxial full-shell Al-InAs NWs. Two device layouts are used, one with the S-island contacted directly (type-I devices) and the other with the S-island placed between the InAs segments (type-II devices). We demonstrate the suppression of superconductivity by critical current, critical voltage or critical Joule power, as determined by the superconducting or normal state of the reservoirs and the quasiparticle dwell time in the island. Our experiments illuminate potential shortcomings of transport spectroscopy in floating S-islands related to non-equilibrium superconductivity. 


Samples used in this study are fabricated from nominally identical InAs NWs grown by molecular beam epitaxy, with an in-situ deposited Al shell fully surrounding the NW. A scanning electron micrograph of the as grown NW array is given in Fig.~\ref{Fig1}a with further growth details provided in  Supplemental Materials (SM~\cite{supplemental}). Individual NWs are dry-transferred with a home-made micro-manipulator onto pre-patterned $\sim150$\,nm thick Au pads, which serve to align and suspend NWs above the $\mathrm{Si/SiO_2}$ substrate. Transport and noise measurements are performed in a quasi-four point setup in a $^3$He cryostat at base temperatures of $T_\mathrm{0}\approx\,$0.45-0.5\,K with the sample immersed in liquid. Altogether we studied two type-I devices and five type-II devices with very similar results among each group. 
\begin{figure}[t!]
\begin{center}
\vspace{0mm}
  \includegraphics[width=1\linewidth]{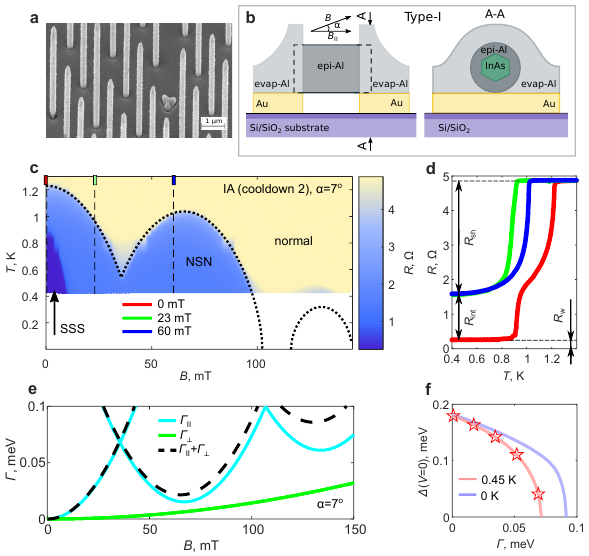}
\end{center}
  \caption{{Superconductivity in equilibrium in type-I devices.} a: scanning electron micrograph of the as-grown Al-InAs NW array. b:
schematic layout of a type-I device and cross-section in the contact region. c: color-scale plot of $R(B,T)$ in device IA (cooldown 2). Dotted line is the fit to $T_\mathrm{c}(B)$ of the epi-Al. d: $R(T)$ traces at fixed $B$-fields, corresponding to 
the vertical dashed lines in panel c. e: depairing factors $\Gamma$, $\Gamma_\parallel$ and $\Gamma_\perp$ as a function of the $B$-field, calculated to fit the data in panel c. f: superconducting order parameter in dependence of $\Gamma$ at two different temperatures (see legend). Discrete points $\Delta(\Gamma)$ marked by stars are the same as in Fig.~\ref{Fig2}d.} 
	\label{Fig1}
\end{figure}

We start from superconducting properties in equilibrium, characterized in type-I devices. Here, ohmic contacts are established directly to the shell, see the sample layout and contact cross-section in Fig.~\ref{Fig1}b. High quality interface between epitaxial aluminum (epi-Al) and $250\,$nm thick e-gun evaporated aluminum (evap-Al) is achieved via in-situ Ar milling. Fig.~\ref{Fig1}c is a color-scale plot of the linear response resistance ($R$) in device IA as a function of $B$ and $T$ (cooldown 2). Three regimes are identified: normal high-$T$ regime, superconducting low-$T$ and low-$B$ regime (SSS) and intermediate regime with the superconducting  shell and normal contacts (NSN). The regimes change at the transitions of the epi-Al and evap-Al from the normal to the superconducting state, with representative $R(T)$ curves displayed in Fig.~\ref{Fig1}d. In $B$=$0$ (red line) $R(T)$ exhibits two steps at the critical temperatures ($T_\mathrm{c}$) of the epi-Al ($T_\mathrm{c}^0\approx1.23\,$K)  and evap-Al ($T_\mathrm{c}\approx0.95\,$K). In the latter case, the reduced $T_\mathrm{c}$ is a result of inverse proximity effect from the Au layer. At lower $T$ the resistance saturates at $R_\mathrm{w}\approx0.26\,\mathrm{\Omega}$, which we attribute to the wiring contribution. Two other traces taken above the critical $B$-field of the evap-Al show only a single step on the $R(T)$ at the $T_\mathrm{c}$ of the epi-Al. This data gives the resistances of evap-Al/epi-Al interfaces $R_\mathrm{int}\approx1.3\,\mathrm{\Omega}$ and of the epi-Al shell $R_\mathrm{sh}\approx3.3\,\mathrm{\Omega}$. As shown in the SM, a series contribution of the contact pads in $R_\mathrm{int}$ is negligible.

The $T_\mathrm{c}$ of the epi-Al in Fig.~\ref{Fig1}c exhibits the Little-Parks (LP) oscillations in the $B$-field, which enables to extract microscopic parameters of the shell. The dependence of $T_\mathrm{c}(B)$ is controlled by the depairing factor $\Gamma=\Gamma_\parallel+\Gamma_\perp$, which has contributions from parallel ($B_\parallel\approx B$) and perpendicular ($B_\perp\approx \alpha B$) components of the $B$-field. Here $\alpha$ is a small angle between the NW axis and the $B$-field, which is not controlled in the experiment and treated as a fit parameter, separately in each device and in each cooldown. In our calculations we closely follow the Usadel theory in the formalism of Ref.~\cite{anthore2003}, see the SM for the details. The $T_\mathrm{c}(B)$ is found from the Abrikosov-Gorkov equation~\cite{vaitiekenas2020,ibabe2023} (dotted line in Fig.~\ref{Fig1}c). $\Gamma$ is derived in the approximation of a cylindrical shell with the inner radius of $\rho_\mathrm{i}$ and thickness of $t$, without the assumption that $t\ll\rho_\mathrm{i}$. The best fits provide $\rho_\mathrm{i}$=80$\pm$4\,nm, $t$=42\,nm and diffusion coefficient $D$=69\,$\text{cm}^2/\text{s}$. The order parameter $\Delta_0\approx187\,\mu\mathrm{eV}$ and the superconducting coherence length $\xi_0\equiv\sqrt{\hbar D/\Delta_0} \approx 156\,\mathrm{nm}$ in the limit of $B$=$0$, $T$=$0$ are obtained from the BCS relation ${\Delta_0\approx1.76 k_\mathrm{B}T_\mathrm{c}^0}$. The calculated dependencies  
$\Gamma(B)$ in device IA (cooldown 2) in different LP lobes and $\Delta(\Gamma)$ for $T$=$T_\mathrm{0}$ and $T$=0 are plotted, respectively, in Figs.~\ref{Fig1}e and~\ref{Fig1}f. 

Next we investigate the fate of shell superconductivity in response to transport current ($I$). Here, three different scenarios can be expected. In devices with superconducting reservoirs directly contacting the shell, the superconductivity breaks down in a conventional way at the shell critical current $I_\mathrm{c}$. This is realized in type-I devices in the SSS regime. In devices with normal reservoirs, the resistance is always finite and quasiparticle non-equilibrium plays crucial role~\cite{keizer2006}. The electronic energy distribution (EED, $f(\varepsilon)$) is then determined by a competition of finite $V$ and energy relaxation. Without relaxation, $f(\varepsilon)$ is a non-equilibrium double-step $f_\mathrm{NEQ}(V) = \left[f_0(\varepsilon-V/2,T_\mathrm{0})+f_0(\varepsilon+V/2,T_\mathrm{0})\right]/2$, where $f_0(\varepsilon,T)$ is the Fermi-Dirac EED at a given $T$. This EED implies symmetric coupling to the reservoirs~\cite{keizer2006,snyman2009}, that agrees with the experimental data. For strong relaxation, local equilibrium is achieved with $f_\mathrm{LEQ}(T_\mathrm{e})=f_0(\varepsilon,T_\mathrm{e})$, where $T_\mathrm{e}>T_\mathrm{0}$ is the electronic temperature in the island. In the first case the superconductivity is destroyed at a critical voltage~\cite{keizer2006,snyman2009,vercruyssen2012,bubis2021} $|V_\mathrm{C}|\sim\Delta/e$, whereas in the second case it collapses at $T_\mathrm{e}$=$T_\mathrm{c}$. The two limiting cases are realized, respectively, in type-I and type-II devices.  

\begin{figure}[t!]
\begin{center}
\vspace{0mm}
  \includegraphics[width=1\linewidth]{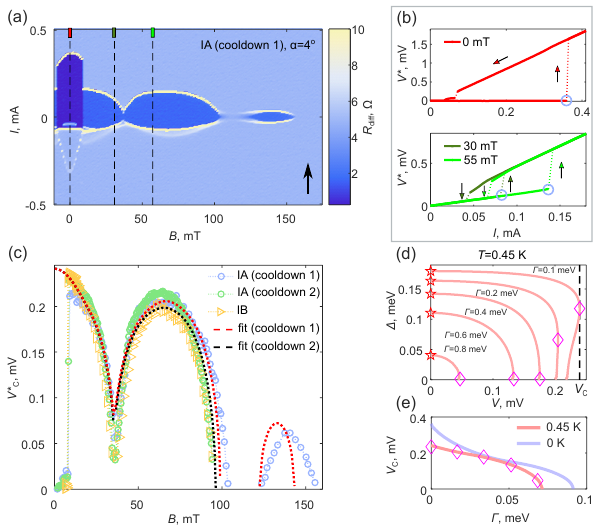}
\end{center}
  \caption{Non-equilibrium in type-I devices. a: color-scale plot of $R_\mathrm{diff}(B,I)$ in the device IA (cooldown 1) for current sweep direction indicated by an arrow. b: $I$-$V^*$ curves taken at fixed $B$-field values in the SSS and NSN regimes for both current sweep directions (arrows). The $B$-fields are indicated in legends and correspond to the vertical dashed lines in panel a. c: critical voltage extracted from the $I$-$V^*$ curves for the device IA in two cooldowns and device IB (symbols). Dashed lines are fits to the data in device IA. d: calculated dependencies $\Delta(V)$ for a set of $\Gamma$, used to obtain the equilibrium order parameter (stars) and critical voltage (diamonds). Dashed line represents $V_\mathrm{C}$ for $\Gamma=0.1\,$meV.  e: Calculated $V_\mathrm{C}(\Gamma)$ at $T$=$T_0$ and $T$=$0$, reproducing the data from panel d (diamonds).} 
	\label{Fig2}
\end{figure}
Fig.~\ref{Fig2} summarizes the non-equilibrium response in type-I device IA (cooldown 1). Fig.~\ref{Fig2}a is a color-scale plot  of the differential resistance $R_\mathrm{diff} = dV^*/dI$  at $T$=$T_0$ in dependence on $B$ and $I$, with the current sweep direction indicated by the arrow. Here, $V^*\equiv V-IR_\mathrm{w}$ is the actual bias on the device with subtracted wiring contribution. Vertical dashed lines correspond to  $I$-$V^*$ curves displayed in Fig.~\ref{Fig2}b for both sweep directions. A superconducting behavior with $V^*\approx0$ is found in the SSS regime (upper panel), whereas finite-resistance superconductivity is evident in the NSN regime (lower panel). In both cases, the usual huge hysteresis is found~\cite{vercruyssen2012}. We are interested in a suppression of the superconducting state, which occurs at increasing $|V^*|$ and is manifested by a single jump on the $I$-$V^*$ curves. Critical voltages $V^*_\mathrm{C}$ measured right before this jump are exemplified by circles in Fig.~\ref{Fig2}b. Symbols in Fig.~\ref{Fig2}c display the $B$-field evolution of $V^*_\mathrm{C}$ in device IA (two cooldowns) and in device IB. In the SSS regime, a small residual voltage is measured, possibly originating from phase-slips or vortices in $B_\perp\neq0$, thus the superconductivity is destroyed in a conventional way at a critical current. The value of $I_\mathrm{c}(B=0)\approx0.35$\,mA is a factor of two smaller compared to the thermodynamical critical current of the epi-Al, the difference most likely coming from the interface resistance (see the SM). By contrast, in the NSN regime the superconductivity collapses at smaller $I$ and at $V^*_\mathrm{C}\sim\Delta/e$. The measured $V^*_\mathrm{C}(B)$ is consistent among the devices and cooldowns, with deviations at higher $B$-fields caused by variations of $\alpha$.   
 
 We explain the evolution of $V^*_\mathrm{C}(B)$ in the NSN regime by the Usadel theory, taking into account the non-equilibrium EED in spirit of Ref.~\cite{keizer2006}. We find the solution in the depth of the S-island, at distances larger than $\xi_0$ from the ends, where the charge-imbalance decays and the non-equilibrium EED is of the form $f_\mathrm{NEQ}(V)$. Self-consistent numerical procedure to find $\Delta(V)$ is detailed in the SM. Fig.~\ref{Fig2}d shows the results for a set of $\Gamma$ (solid lines). Data in equilibrium ($V$=$0$) is the same as in Fig.~\ref{Fig1}f (stars). At increasing $V$, $\Delta$ gets suppressed, so that no solution exists above certain $V_\mathrm{C}$ (diamonds, dashed line). At small $\Gamma$ the solution is bistable just below $V_\mathrm{C}$, consistent with previous results~\cite{keizer2006,snyman2009,vercruyssen2012,bubis2021}.  Calculated $V_\mathrm{C}(\Gamma)$ is shown in Fig.~\ref{Fig2}e along with the symbols from Fig.~\ref{Fig2}d. Using the $\Gamma$-$B$ correspondence the dependencies $V_\mathrm{C}(B)$ are obtained and plotted in Fig.~\ref{Fig2}c for the device IA in two cooldowns (dashed lines). Near perfect agreement with the experiment ensures that the relaxation-free scenario of non-equilibrium superconductivity is realized in type-I devices.  
\begin{figure}[t]
\begin{center}
\vspace{0mm}
  \includegraphics[width=1\linewidth]{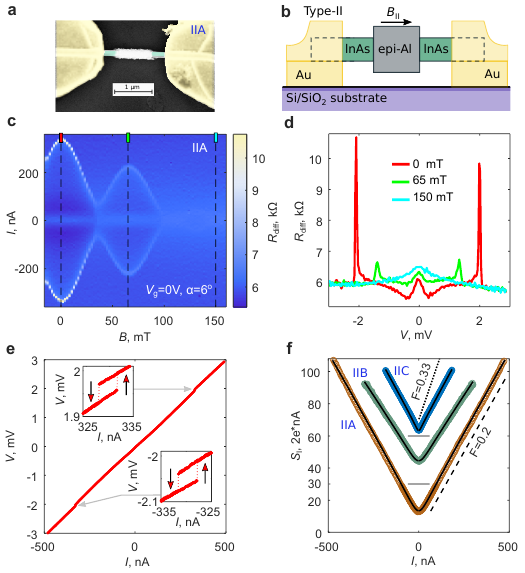}
\end{center}
  \caption{Non-equilibrium in type-II devices. a,b: electron micrograph of the device IIA (false colors) and schematic layout of type-II devices. c: color-scale plot of $R_\mathrm{diff}(B,I)$ in device IIA. d: bias dependencies of $R_\mathrm{diff}$ at fixed $B$-fields, corresponding to the dashed lines in panel c. e: $I$-$V$ curve measured at $B=0$, showing tiny voltage jumps at high biases $V\approx2$\,mV. Insets magnify these features, demonstrating weak hysteresis with respect to current sweep direction. f:
noise spectral density in devices IIA, IIB and IIC as a function of $I$ in  the normal transport regime at high enough $B$ (symbols). Two upper traces are shifted upward for clarity, with zero levels marked by thin solid lines. Dashed and dotted guide lines have slopes corresponding to Fano factor of $F$=0.2 and $F_\mathrm{D}$ = 1/3, respectively. Solid lines are fits taking into account the relaxation in the S-island (see the SM).} 
	\label{Fig3}
\end{figure}
\begin{figure}[t!]
\begin{center}
\vspace{0mm}
  \includegraphics[width=1\linewidth]{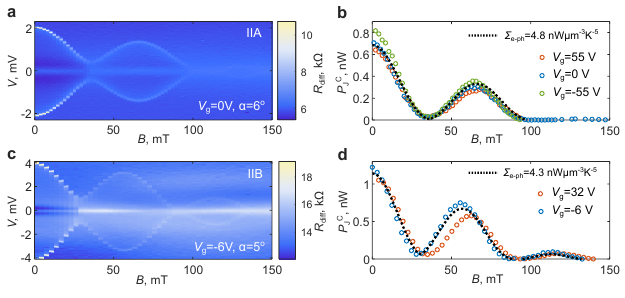}
\end{center}
  \caption{Critical Joule power in type-II devices. a: color-scale plot of $R_\mathrm{diff}(B,V)$ in device IIA ($V_\mathrm{g}$ and $\alpha$ indicated in the legend). b:  $B$-dependencies of the critical Joule power for the same device (symbols) and the fit to the e-ph relaxation model (dashed line). Gate voltages and the e-ph cooling power are indicated in the legend. c, d: the same in device IIB.} 
	\label{Fig4}
\end{figure}

We now switch to type-II devices, where the selectively etched shell forms a floating S-island of the length $L=\text{0.5 - 2}\,\mu\mathrm{m}$, see the micrograph and sketch in Figs.~\ref{Fig3}a and~\ref{Fig3}b. Ohmic contacts are defined via evaporation of Cr/Au with ex-situ  passivation of the native oxide in ammonium polysulfide~\cite{sourribes2013,paajaste2015,gul2017}. Thanks to the InAs segments, 
the device resistance is about four orders of magnitude higher than $R_\mathrm{int}$ in type-I devices and is controllable by the back gate voltage $V_\mathrm{g}$. Figs.~\ref{Fig3}c and~\ref{Fig3}d show, respectively, the color-scale plot $R_\mathrm{diff}(I,B)$ in device IIA and representative traces $R_\mathrm{diff}(V)$ at fixed $B$, corresponding to vertical dashed lines in the color-plot. Low bias behavior of $R_\mathrm{diff}$ is a combination of the superconducting proximity effect in diffusive NS junctions with non-ideal interface~\cite{bubis2017,denisov2021,denisov2022}, observable at low $B$-fields, and Coulomb effects~\cite{nazarov1999,golubev2001,yeyati2001}, which contribute a broad zero-bias resistance maximum, observable at  high $B$-fields. We do not discuss these device specific and $V_\mathrm{g}$-dependent properties and concentrate on a sharp resistance peak observed in all devices at much higher $V$. The LP oscillations of the peak position (Fig.~\ref{Fig3}c), and  the tiny voltage jump the peak originates from (Fig.~\ref{Fig3}e), show that this feature is associated with the collapse of superconductivity. Corresponding voltage jumps are most pronounced at $B$=$0$ and demonstrate weak hysteresis, see the insets of Fig.~\ref{Fig3}e. At increasing $B$ they smear out and the visibility of the LP oscillations reduces (Fig.~\ref{Fig3}c). 

The observation of superconducting state at voltages $|V|\gg\Delta/e$ implies strong energy relaxation. Otherwise, as found in type-I devices, the double-step EED $f_\mathrm{NEQ}(V)$ in the S-island would not be compatible with the superconductivity. Although in the type-II devices a moderate asymmetry of the couplings to the reservoirs can affect the EED and reduce the effect of non-equilibrium~\cite{snyman2009,bubis2021}, it is too weak to maintain the superconducting state at mV-range biases. A direct test of the relaxation is achieved via shot noise measurements in the normal state, shown in Fig.~\ref{Fig3}f, see SM for the details. The noise spectral density (symbols) exhibits a shot noise behavior $S_\mathrm{I}\approx 2eFI$ with Fano factors $F\approx0.2$ (dashed line). This value is considerably reduced compared to the universal $F_\mathrm{D}=1/3$ in diffusive conductors without relaxation~\cite{nagaev1992,beenakker1992}, usually found in InAs NWs~\cite{tikhonov2016a,tikhonov2016b,denisov2022} (dotted line). The reduction of $F$ is a result of strong e-ph relaxation in the S-island~\cite{dejong1996}. We assume local equilibrium EED $f_\mathrm{LEQ}(T_\mathrm{e})$ with the  electronic temperature, which obeys the heat balance equation $\frac{1}{2}P_\mathrm{J} = \mathcal{V}_\mathrm{Al} \Sigma_\mathrm{e-ph}\left(T_\mathrm{e}^5-T_\mathrm{0}^5\right)$. Here, $P_\mathrm{J}\equiv IV$ is the Joule power released in the semiconducting segments, half of which is dissipated in the S-island, $\mathcal{V}_\mathrm{Al}$ is the volume of the epi-Al and $\Sigma_\mathrm{e-ph}$ is the e-ph cooling power. Note a conceptual difference from Ref.~\cite{ibabe2023}, in which the Joule power flows into the reservoirs by the electronic heat conduction, while the e-ph relaxation is neglected. The above equation simultaneously explains the shot noise in Fig.~\ref{Fig3}f (solid lines) and the suppression of superconductivity by transport current. In the latter case, the superconductivity collapses at the critical Joule power ($P_\mathrm{J}^\mathrm{c}$), which corresponds to $T_\mathrm{e}$=$T_\mathrm{c}(B)$. The dependencies $P_\mathrm{J}^\mathrm{c}(B)$ in devices IIA and IIB are shown in Figs.~\ref{Fig4}b and~\ref{Fig4}d (symbols) and correspond, respectively, to the color-plots in Figs.~\ref{Fig4}a and~\ref{Fig4}c. We fit the data in these and other devices with similar $\Sigma_\mathrm{e-ph}\approx4.8\pm1\,\text{nW}\mu\text{m}^{-3}\text{K}^{-5}$, see the dashed lines (more data in the SM). This value corresponds to the e-ph relaxation time of $\tau_\mathrm{e-ph}\approx 60\,$ns at $T$=$T_0$, in agreement with independent measurements in aluminum~\cite{pinsolle2016}.

The origin of different behavior in type-I and type-II devices is in the ratio of quasiparticle dwell time in the S-island $\tau_\mathrm{dwell}$, controlled by the coupling to the reservoirs, and $\tau_\mathrm{e-ph}$. Type-I devices are strongly coupled to reservoirs and the dwell time is mainly limited by diffusion $\tau_\mathrm{dwell}^\mathrm{I}\sim L^2/D\approx1\,$ns for the typical $L\approx3\,\mu\mathrm{m}$. In type-II devices $L$ is smaller, however, the coupling to reservoirs is very weak owing to highly resistive InAs segments. Thus, the dwell time is renormalized  by the ratio of the numbers of the eigenmodes in epi-Al and in InAs or, roughly, by the ratio of semiconductor and superconductor resistances in the normal state ($\sim10^4$), giving $\tau_\mathrm{dwell}^\mathrm{II}\sim 1\,\mu\mathrm{s}$. The relation $\tau_\mathrm{dwell}^\mathrm{I}\ll\tau_\mathrm{e-ph}\ll\tau_\mathrm{dwell}^\mathrm{II}$ naturally explains the relaxation-free regime in type-I devices and strong relaxation regime in type-II devices. Semi-super research usually deals with superconducting islands analogous to our type-II devices~\cite{fu2010,ulrich2015,albrecht2016,vayrynen2021,lai2021,souto2022,valentini2022}. It is illuminating to discuss how such systems could fall in the non-equilibrium regime observed in type-I devices. Sizable decrease of the $\tau_\mathrm{dwell}$ by populating more conduction channels is not feasible in semiconducting NWs. Figs.~\ref{Fig4}b and~\ref{Fig4}d demonstrate a weak $V_\mathrm{g}$ dependence of $P_\mathrm{J}^\mathrm{c}$, indicating only minor deviations from local equilibrium in our experiment. Note, however, that lowering the temperature leads to the increase of $\tau_{\text{e-ph}}\propto T^{-3}$ so that energy relaxation slows down significantly. Critical voltages on the order of $\Delta/e$ found at mK temperatures in the recent study~\cite{ibabe2023a} may indicate the relevance of non-equilibrium.

In summary, our results illuminate the way in which the superconducting order parameter and the bias and relaxation controlled quasiparticle population are bound with each other in floating S-islands. This binding indicates a general shortcoming of the transport spectroscopy in semi-super hybrids, since the quasiparticle excitation spectrum becomes dependent on the bias voltage and relaxation. 

We acknowledge valuable advices of Ya.V. Fominov and A.S. Mel'nikov on the Usadel theory and fabrication help of S.V. Egorov. We thank A.V. Bubis for his input on the early stages of this work and for useful remarks. VSK is grateful to T.M. Klapwijk for the early illuminating discussions of non-equilibrium superconductivity. The work was supported by the Russian Science Foundation project 22-12-00342.

\nocite{ruhstorfer2021,delgiudice2020,rudolph2013,tikhonov2014a}


\begin{thebibliography}{77}%
\makeatletter
\providecommand \@ifxundefined [1]{%
 \@ifx{#1\undefined}
}%
\providecommand \@ifnum [1]{%
 \ifnum #1\expandafter \@firstoftwo
 \else \expandafter \@secondoftwo
 \fi
}%
\providecommand \@ifx [1]{%
 \ifx #1\expandafter \@firstoftwo
 \else \expandafter \@secondoftwo
 \fi
}%
\providecommand \natexlab [1]{#1}%
\providecommand \enquote  [1]{``#1''}%
\providecommand \bibnamefont  [1]{#1}%
\providecommand \bibfnamefont [1]{#1}%
\providecommand \citenamefont [1]{#1}%
\providecommand \href@noop [0]{\@secondoftwo}%
\providecommand \href [0]{\begingroup \@sanitize@url \@href}%
\providecommand \@href[1]{\@@startlink{#1}\@@href}%
\providecommand \@@href[1]{\endgroup#1\@@endlink}%
\providecommand \@sanitize@url [0]{\catcode `\\12\catcode `\$12\catcode
  `\&12\catcode `\#12\catcode `\^12\catcode `\_12\catcode `\%12\relax}%
\providecommand \@@startlink[1]{}%
\providecommand \@@endlink[0]{}%
\providecommand \url  [0]{\begingroup\@sanitize@url \@url }%
\providecommand \@url [1]{\endgroup\@href {#1}{\urlprefix }}%
\providecommand \urlprefix  [0]{URL }%
\providecommand \Eprint [0]{\href }%
\providecommand \doibase [0]{https://doi.org/}%
\providecommand \selectlanguage [0]{\@gobble}%
\providecommand \bibinfo  [0]{\@secondoftwo}%
\providecommand \bibfield  [0]{\@secondoftwo}%
\providecommand \translation [1]{[#1]}%
\providecommand \BibitemOpen [0]{}%
\providecommand \bibitemStop [0]{}%
\providecommand \bibitemNoStop [0]{.\EOS\space}%
\providecommand \EOS [0]{\spacefactor3000\relax}%
\providecommand \BibitemShut  [1]{\csname bibitem#1\endcsname}%
\let\auto@bib@innerbib\@empty
\bibitem [{\citenamefont {Lutchyn}\ \emph {et~al.}(2010)\citenamefont
  {Lutchyn}, \citenamefont {Sau},\ and\ \citenamefont
  {Das~Sarma}}]{lutchyn2010}%
  \BibitemOpen
  \bibfield  {author} {\bibinfo {author} {\bibfnamefont {R.~M.}\ \bibnamefont
  {Lutchyn}}, \bibinfo {author} {\bibfnamefont {J.~D.}\ \bibnamefont {Sau}},\
  and\ \bibinfo {author} {\bibfnamefont {S.}~\bibnamefont {Das~Sarma}},\
  }\bibfield  {title} {\bibinfo {title} {Majorana {{Fermions}} and a
  {{Topological Phase Transition}} in {{Semiconductor-Superconductor
  Heterostructures}}},\ }\href {https://doi.org/10.1103/PhysRevLett.105.077001}
  {\bibfield  {journal} {\bibinfo  {journal} {Physical Review Letters}\
  }\textbf {\bibinfo {volume} {105}},\ \bibinfo {pages} {077001} (\bibinfo
  {year} {2010})}\BibitemShut {NoStop}%
\bibitem [{\citenamefont {Oreg}\ \emph {et~al.}(2010)\citenamefont {Oreg},
  \citenamefont {Refael},\ and\ \citenamefont {Von~Oppen}}]{oreg2010}%
  \BibitemOpen
  \bibfield  {author} {\bibinfo {author} {\bibfnamefont {Y.}~\bibnamefont
  {Oreg}}, \bibinfo {author} {\bibfnamefont {G.}~\bibnamefont {Refael}},\ and\
  \bibinfo {author} {\bibfnamefont {F.}~\bibnamefont {Von~Oppen}},\ }\bibfield
  {title} {\bibinfo {title} {Helical {{Liquids}} and {{Majorana Bound States}}
  in {{Quantum Wires}}},\ }\href
  {https://doi.org/10.1103/PhysRevLett.105.177002} {\bibfield  {journal}
  {\bibinfo  {journal} {Physical Review Letters}\ }\textbf {\bibinfo {volume}
  {105}},\ \bibinfo {pages} {177002} (\bibinfo {year} {2010})}\BibitemShut
  {NoStop}%
\bibitem [{\citenamefont {Prada}\ \emph {et~al.}(2020)\citenamefont {Prada},
  \citenamefont {{San-Jose}}, \citenamefont {De~Moor}, \citenamefont {Geresdi},
  \citenamefont {Lee}, \citenamefont {Klinovaja}, \citenamefont {Loss},
  \citenamefont {Nyg{\aa}rd}, \citenamefont {Aguado},\ and\ \citenamefont
  {Kouwenhoven}}]{prada2020}%
  \BibitemOpen
  \bibfield  {author} {\bibinfo {author} {\bibfnamefont {E.}~\bibnamefont
  {Prada}}, \bibinfo {author} {\bibfnamefont {P.}~\bibnamefont {{San-Jose}}},
  \bibinfo {author} {\bibfnamefont {M.~W.~A.}\ \bibnamefont {De~Moor}},
  \bibinfo {author} {\bibfnamefont {A.}~\bibnamefont {Geresdi}}, \bibinfo
  {author} {\bibfnamefont {E.~J.~H.}\ \bibnamefont {Lee}}, \bibinfo {author}
  {\bibfnamefont {J.}~\bibnamefont {Klinovaja}}, \bibinfo {author}
  {\bibfnamefont {D.}~\bibnamefont {Loss}}, \bibinfo {author} {\bibfnamefont
  {J.}~\bibnamefont {Nyg{\aa}rd}}, \bibinfo {author} {\bibfnamefont
  {R.}~\bibnamefont {Aguado}},\ and\ \bibinfo {author} {\bibfnamefont {L.~P.}\
  \bibnamefont {Kouwenhoven}},\ }\bibfield  {title} {\bibinfo {title} {From
  {{Andreev}} to {{Majorana}} bound states in hybrid
  superconductor{\textendash}semiconductor nanowires},\ }\href
  {https://doi.org/10.1038/s42254-020-0228-y} {\bibfield  {journal} {\bibinfo
  {journal} {Nature Reviews Physics}\ }\textbf {\bibinfo {volume} {2}},\
  \bibinfo {pages} {575} (\bibinfo {year} {2020})}\BibitemShut {NoStop}%
\bibitem [{\citenamefont {Chang}\ \emph {et~al.}(2015)\citenamefont {Chang},
  \citenamefont {Albrecht}, \citenamefont {Jespersen}, \citenamefont
  {Kuemmeth}, \citenamefont {Krogstrup}, \citenamefont {Nyg{\aa}rd},\ and\
  \citenamefont {Marcus}}]{chang2015}%
  \BibitemOpen
  \bibfield  {author} {\bibinfo {author} {\bibfnamefont {W.}~\bibnamefont
  {Chang}}, \bibinfo {author} {\bibfnamefont {S.~M.}\ \bibnamefont {Albrecht}},
  \bibinfo {author} {\bibfnamefont {T.~S.}\ \bibnamefont {Jespersen}}, \bibinfo
  {author} {\bibfnamefont {F.}~\bibnamefont {Kuemmeth}}, \bibinfo {author}
  {\bibfnamefont {P.}~\bibnamefont {Krogstrup}}, \bibinfo {author}
  {\bibfnamefont {J.}~\bibnamefont {Nyg{\aa}rd}},\ and\ \bibinfo {author}
  {\bibfnamefont {C.~M.}\ \bibnamefont {Marcus}},\ }\bibfield  {title}
  {\bibinfo {title} {Hard gap in epitaxial
  semiconductor{\textendash}superconductor nanowires},\ }\href
  {https://doi.org/10.1038/nnano.2014.306} {\bibfield  {journal} {\bibinfo
  {journal} {Nature Nanotechnology}\ }\textbf {\bibinfo {volume} {10}},\
  \bibinfo {pages} {232} (\bibinfo {year} {2015})}\BibitemShut {NoStop}%
\bibitem [{\citenamefont {Krogstrup}\ \emph {et~al.}(2015)\citenamefont
  {Krogstrup}, \citenamefont {Ziino}, \citenamefont {Chang}, \citenamefont
  {Albrecht}, \citenamefont {Madsen}, \citenamefont {Johnson}, \citenamefont
  {Nyg{\aa}rd}, \citenamefont {Marcus},\ and\ \citenamefont
  {Jespersen}}]{krogstrup2015}%
  \BibitemOpen
  \bibfield  {author} {\bibinfo {author} {\bibfnamefont {P.}~\bibnamefont
  {Krogstrup}}, \bibinfo {author} {\bibfnamefont {N.~L.~B.}\ \bibnamefont
  {Ziino}}, \bibinfo {author} {\bibfnamefont {W.}~\bibnamefont {Chang}},
  \bibinfo {author} {\bibfnamefont {S.~M.}\ \bibnamefont {Albrecht}}, \bibinfo
  {author} {\bibfnamefont {M.~H.}\ \bibnamefont {Madsen}}, \bibinfo {author}
  {\bibfnamefont {E.}~\bibnamefont {Johnson}}, \bibinfo {author} {\bibfnamefont
  {J.}~\bibnamefont {Nyg{\aa}rd}}, \bibinfo {author} {\bibfnamefont {C.~M.}\
  \bibnamefont {Marcus}},\ and\ \bibinfo {author} {\bibfnamefont {T.~S.}\
  \bibnamefont {Jespersen}},\ }\bibfield  {title} {\bibinfo {title} {Epitaxy of
  semiconductor{\textendash}superconductor nanowires},\ }\href
  {https://doi.org/10.1038/nmat4176} {\bibfield  {journal} {\bibinfo  {journal}
  {Nature Materials}\ }\textbf {\bibinfo {volume} {14}},\ \bibinfo {pages}
  {400} (\bibinfo {year} {2015})}\BibitemShut {NoStop}%
\bibitem [{\citenamefont {G{\"u}l}\ \emph {et~al.}(2017)\citenamefont
  {G{\"u}l}, \citenamefont {Zhang}, \citenamefont {{de Vries}}, \citenamefont
  {{van Veen}}, \citenamefont {Zuo}, \citenamefont {Mourik}, \citenamefont
  {{Conesa-Boj}}, \citenamefont {Nowak}, \citenamefont {{van Woerkom}},
  \citenamefont {{Quintero-P{\'e}rez}}, \citenamefont {Cassidy}, \citenamefont
  {Geresdi}, \citenamefont {Koelling}, \citenamefont {Car}, \citenamefont
  {Plissard}, \citenamefont {Bakkers},\ and\ \citenamefont
  {Kouwenhoven}}]{gul2017}%
  \BibitemOpen
  \bibfield  {author} {\bibinfo {author} {\bibfnamefont {{\"O}.}~\bibnamefont
  {G{\"u}l}}, \bibinfo {author} {\bibfnamefont {H.}~\bibnamefont {Zhang}},
  \bibinfo {author} {\bibfnamefont {F.~K.}\ \bibnamefont {{de Vries}}},
  \bibinfo {author} {\bibfnamefont {J.}~\bibnamefont {{van Veen}}}, \bibinfo
  {author} {\bibfnamefont {K.}~\bibnamefont {Zuo}}, \bibinfo {author}
  {\bibfnamefont {V.}~\bibnamefont {Mourik}}, \bibinfo {author} {\bibfnamefont
  {S.}~\bibnamefont {{Conesa-Boj}}}, \bibinfo {author} {\bibfnamefont {M.~P.}\
  \bibnamefont {Nowak}}, \bibinfo {author} {\bibfnamefont {D.~J.}\ \bibnamefont
  {{van Woerkom}}}, \bibinfo {author} {\bibfnamefont {M.}~\bibnamefont
  {{Quintero-P{\'e}rez}}}, \bibinfo {author} {\bibfnamefont {M.~C.}\
  \bibnamefont {Cassidy}}, \bibinfo {author} {\bibfnamefont {A.}~\bibnamefont
  {Geresdi}}, \bibinfo {author} {\bibfnamefont {S.}~\bibnamefont {Koelling}},
  \bibinfo {author} {\bibfnamefont {D.}~\bibnamefont {Car}}, \bibinfo {author}
  {\bibfnamefont {S.~R.}\ \bibnamefont {Plissard}}, \bibinfo {author}
  {\bibfnamefont {E.~P. A.~M.}\ \bibnamefont {Bakkers}},\ and\ \bibinfo
  {author} {\bibfnamefont {L.~P.}\ \bibnamefont {Kouwenhoven}},\ }\bibfield
  {title} {\bibinfo {title} {Hard superconducting gap in {{InSb}} nanowires},\
  }\href {https://doi.org/10.1021/acs.nanolett.7b00540} {\bibfield  {journal}
  {\bibinfo  {journal} {Nano Letters}\ }\textbf {\bibinfo {volume} {17}},\
  \bibinfo {pages} {2690} (\bibinfo {year} {2017})}\BibitemShut {NoStop}%
\bibitem [{\citenamefont {Bubis}(2017)}]{bubis2017}%
  \BibitemOpen
  \bibfield  {author} {\bibinfo {author} {\bibfnamefont {A.~V.}\ \bibnamefont
  {Bubis}},\ }\bibfield  {title} {\bibinfo {title} {Proximity effect and
  interface transparency in {{Al}}/{{InAs-nanowire}}/{{Al}} diffusive
  junctions},\ }\href {https://doi.org/10.1088/1361-6641/aa7eef} {\bibfield
  {journal} {\bibinfo  {journal} {Semicond. Sci. Technol.}\ }\textbf {\bibinfo
  {volume} {32}},\ \bibinfo {pages} {094007} (\bibinfo {year}
  {2017})}\BibitemShut {NoStop}%
\bibitem [{\citenamefont {J{\"u}nger}\ \emph {et~al.}(2020)\citenamefont
  {J{\"u}nger}, \citenamefont {Delagrange}, \citenamefont {Chevallier},
  \citenamefont {Lehmann}, \citenamefont {Dick}, \citenamefont {Thelander},
  \citenamefont {Klinovaja}, \citenamefont {Loss}, \citenamefont
  {Baumgartner},\ and\ \citenamefont {Sch{\"o}nenberger}}]{junger2020}%
  \BibitemOpen
  \bibfield  {author} {\bibinfo {author} {\bibfnamefont {C.}~\bibnamefont
  {J{\"u}nger}}, \bibinfo {author} {\bibfnamefont {R.}~\bibnamefont
  {Delagrange}}, \bibinfo {author} {\bibfnamefont {D.}~\bibnamefont
  {Chevallier}}, \bibinfo {author} {\bibfnamefont {S.}~\bibnamefont {Lehmann}},
  \bibinfo {author} {\bibfnamefont {K.~A.}\ \bibnamefont {Dick}}, \bibinfo
  {author} {\bibfnamefont {C.}~\bibnamefont {Thelander}}, \bibinfo {author}
  {\bibfnamefont {J.}~\bibnamefont {Klinovaja}}, \bibinfo {author}
  {\bibfnamefont {D.}~\bibnamefont {Loss}}, \bibinfo {author} {\bibfnamefont
  {A.}~\bibnamefont {Baumgartner}},\ and\ \bibinfo {author} {\bibfnamefont
  {C.}~\bibnamefont {Sch{\"o}nenberger}},\ }\bibfield  {title} {\bibinfo
  {title} {Magnetic-{{Field-Independent Subgap States}} in {{Hybrid Rashba
  Nanowires}}},\ }\href {https://doi.org/10.1103/PhysRevLett.125.017701}
  {\bibfield  {journal} {\bibinfo  {journal} {Physical Review Letters}\
  }\textbf {\bibinfo {volume} {125}},\ \bibinfo {pages} {017701} (\bibinfo
  {year} {2020})}\BibitemShut {NoStop}%
\bibitem [{\citenamefont {Yu}\ \emph {et~al.}(2023)\citenamefont {Yu},
  \citenamefont {Woods}, \citenamefont {Chen}, \citenamefont {Badawy},
  \citenamefont {Bakkers}, \citenamefont {Stanescu},\ and\ \citenamefont
  {Frolov}}]{yu2023}%
  \BibitemOpen
  \bibfield  {author} {\bibinfo {author} {\bibfnamefont {P.}~\bibnamefont
  {Yu}}, \bibinfo {author} {\bibfnamefont {B.~D.}\ \bibnamefont {Woods}},
  \bibinfo {author} {\bibfnamefont {J.}~\bibnamefont {Chen}}, \bibinfo {author}
  {\bibfnamefont {G.}~\bibnamefont {Badawy}}, \bibinfo {author} {\bibfnamefont
  {E.~P. A.~M.}\ \bibnamefont {Bakkers}}, \bibinfo {author} {\bibfnamefont
  {T.~D.}\ \bibnamefont {Stanescu}},\ and\ \bibinfo {author} {\bibfnamefont
  {S.~M.}\ \bibnamefont {Frolov}},\ }\bibfield  {title} {\bibinfo {title}
  {Delocalized states in three-terminal superconductor-semiconductor nanowire
  devices},\ }\href {https://doi.org/10.21468/SciPostPhys.15.1.005} {\bibfield
  {journal} {\bibinfo  {journal} {SciPost Physics}\ }\textbf {\bibinfo {volume}
  {15}},\ \bibinfo {pages} {005} (\bibinfo {year} {2023})}\BibitemShut
  {NoStop}%
\bibitem [{\citenamefont {J{\"u}nger}\ \emph {et~al.}(2019)\citenamefont
  {J{\"u}nger}, \citenamefont {Baumgartner}, \citenamefont {Delagrange},
  \citenamefont {Chevallier}, \citenamefont {Lehmann}, \citenamefont {Nilsson},
  \citenamefont {Dick}, \citenamefont {Thelander},\ and\ \citenamefont
  {Sch{\"o}nenberger}}]{junger2019}%
  \BibitemOpen
  \bibfield  {author} {\bibinfo {author} {\bibfnamefont {C.}~\bibnamefont
  {J{\"u}nger}}, \bibinfo {author} {\bibfnamefont {A.}~\bibnamefont
  {Baumgartner}}, \bibinfo {author} {\bibfnamefont {R.}~\bibnamefont
  {Delagrange}}, \bibinfo {author} {\bibfnamefont {D.}~\bibnamefont
  {Chevallier}}, \bibinfo {author} {\bibfnamefont {S.}~\bibnamefont {Lehmann}},
  \bibinfo {author} {\bibfnamefont {M.}~\bibnamefont {Nilsson}}, \bibinfo
  {author} {\bibfnamefont {K.~A.}\ \bibnamefont {Dick}}, \bibinfo {author}
  {\bibfnamefont {C.}~\bibnamefont {Thelander}},\ and\ \bibinfo {author}
  {\bibfnamefont {C.}~\bibnamefont {Sch{\"o}nenberger}},\ }\bibfield  {title}
  {\bibinfo {title} {Spectroscopy of the superconducting proximity effect in
  nanowires using integrated quantum dots},\ }\href
  {https://doi.org/10.1038/s42005-019-0162-4} {\bibfield  {journal} {\bibinfo
  {journal} {Communications Physics}\ }\textbf {\bibinfo {volume} {2}},\
  \bibinfo {pages} {76} (\bibinfo {year} {2019})}\BibitemShut {NoStop}%
\bibitem [{\citenamefont {Stanescu}\ and\ \citenamefont
  {Tewari}(2014)}]{stanescu2014}%
  \BibitemOpen
  \bibfield  {author} {\bibinfo {author} {\bibfnamefont {T.~D.}\ \bibnamefont
  {Stanescu}}\ and\ \bibinfo {author} {\bibfnamefont {S.}~\bibnamefont
  {Tewari}},\ }\bibfield  {title} {\bibinfo {title} {Nonlocality of zero-bias
  anomalies in the topologically trivial phase of {{Majorana}} wires},\ }\href
  {https://doi.org/10.1103/PhysRevB.89.220507} {\bibfield  {journal} {\bibinfo
  {journal} {Physical Review B}\ }\textbf {\bibinfo {volume} {89}},\ \bibinfo
  {pages} {220507} (\bibinfo {year} {2014})}\BibitemShut {NoStop}%
\bibitem [{\citenamefont {Rosdahl}\ \emph {et~al.}(2018)\citenamefont
  {Rosdahl}, \citenamefont {Vuik}, \citenamefont {Kjaergaard},\ and\
  \citenamefont {Akhmerov}}]{rosdahl2018}%
  \BibitemOpen
  \bibfield  {author} {\bibinfo {author} {\bibfnamefont {T.~{\"O}.}\
  \bibnamefont {Rosdahl}}, \bibinfo {author} {\bibfnamefont {A.}~\bibnamefont
  {Vuik}}, \bibinfo {author} {\bibfnamefont {M.}~\bibnamefont {Kjaergaard}},\
  and\ \bibinfo {author} {\bibfnamefont {A.~R.}\ \bibnamefont {Akhmerov}},\
  }\bibfield  {title} {\bibinfo {title} {Andreev rectifier: {{A}} nonlocal
  conductance signature of topological phase transitions},\ }\href
  {https://doi.org/10.1103/PhysRevB.97.045421} {\bibfield  {journal} {\bibinfo
  {journal} {Physical Review B}\ }\textbf {\bibinfo {volume} {97}},\ \bibinfo
  {pages} {045421} (\bibinfo {year} {2018})}\BibitemShut {NoStop}%
\bibitem [{\citenamefont {Lai}\ \emph {et~al.}(2019)\citenamefont {Lai},
  \citenamefont {Sau},\ and\ \citenamefont {Das~Sarma}}]{lai2019}%
  \BibitemOpen
  \bibfield  {author} {\bibinfo {author} {\bibfnamefont {Y.-H.}\ \bibnamefont
  {Lai}}, \bibinfo {author} {\bibfnamefont {J.~D.}\ \bibnamefont {Sau}},\ and\
  \bibinfo {author} {\bibfnamefont {S.}~\bibnamefont {Das~Sarma}},\ }\bibfield
  {title} {\bibinfo {title} {Presence versus absence of end-to-end nonlocal
  conductance correlations in {{Majorana}} nanowires: {{Majorana}} bound states
  versus {{Andreev}} bound states},\ }\href
  {https://doi.org/10.1103/PhysRevB.100.045302} {\bibfield  {journal} {\bibinfo
   {journal} {Physical Review B}\ }\textbf {\bibinfo {volume} {100}},\ \bibinfo
  {pages} {045302} (\bibinfo {year} {2019})}\BibitemShut {NoStop}%
\bibitem [{\citenamefont {M{\'e}nard}\ \emph {et~al.}(2020)\citenamefont
  {M{\'e}nard}, \citenamefont {Anselmetti}, \citenamefont {Martinez},
  \citenamefont {Puglia}, \citenamefont {Malinowski}, \citenamefont {Lee},
  \citenamefont {Choi}, \citenamefont {Pendharkar}, \citenamefont
  {Palmstr{\o}m}, \citenamefont {Flensberg}, \citenamefont {Marcus},
  \citenamefont {Casparis},\ and\ \citenamefont {Higginbotham}}]{menard2020}%
  \BibitemOpen
  \bibfield  {author} {\bibinfo {author} {\bibfnamefont {G.~C.}\ \bibnamefont
  {M{\'e}nard}}, \bibinfo {author} {\bibfnamefont {G.~L.~R.}\ \bibnamefont
  {Anselmetti}}, \bibinfo {author} {\bibfnamefont {E.~A.}\ \bibnamefont
  {Martinez}}, \bibinfo {author} {\bibfnamefont {D.}~\bibnamefont {Puglia}},
  \bibinfo {author} {\bibfnamefont {F.~K.}\ \bibnamefont {Malinowski}},
  \bibinfo {author} {\bibfnamefont {J.~S.}\ \bibnamefont {Lee}}, \bibinfo
  {author} {\bibfnamefont {S.}~\bibnamefont {Choi}}, \bibinfo {author}
  {\bibfnamefont {M.}~\bibnamefont {Pendharkar}}, \bibinfo {author}
  {\bibfnamefont {C.~J.}\ \bibnamefont {Palmstr{\o}m}}, \bibinfo {author}
  {\bibfnamefont {K.}~\bibnamefont {Flensberg}}, \bibinfo {author}
  {\bibfnamefont {C.~M.}\ \bibnamefont {Marcus}}, \bibinfo {author}
  {\bibfnamefont {L.}~\bibnamefont {Casparis}},\ and\ \bibinfo {author}
  {\bibfnamefont {A.~P.}\ \bibnamefont {Higginbotham}},\ }\bibfield  {title}
  {\bibinfo {title} {Conductance-{{Matrix Symmetries}} of a {{Three-Terminal
  Hybrid Device}}},\ }\href {https://doi.org/10.1103/PhysRevLett.124.036802}
  {\bibfield  {journal} {\bibinfo  {journal} {Physical Review Letters}\
  }\textbf {\bibinfo {volume} {124}},\ \bibinfo {pages} {036802} (\bibinfo
  {year} {2020})}\BibitemShut {NoStop}%
\bibitem [{\citenamefont {Denisov}\ \emph {et~al.}(2021)\citenamefont
  {Denisov}, \citenamefont {Bubis}, \citenamefont {Piatrusha}, \citenamefont
  {Titova}, \citenamefont {Nasibulin}, \citenamefont {Becker}, \citenamefont
  {Treu}, \citenamefont {Ruhstorfer}, \citenamefont {Koblm{\"u}ller},
  \citenamefont {Tikhonov},\ and\ \citenamefont {Khrapai}}]{denisov2021}%
  \BibitemOpen
  \bibfield  {author} {\bibinfo {author} {\bibfnamefont {A.~O.}\ \bibnamefont
  {Denisov}}, \bibinfo {author} {\bibfnamefont {A.~V.}\ \bibnamefont {Bubis}},
  \bibinfo {author} {\bibfnamefont {S.~U.}\ \bibnamefont {Piatrusha}}, \bibinfo
  {author} {\bibfnamefont {N.~A.}\ \bibnamefont {Titova}}, \bibinfo {author}
  {\bibfnamefont {A.~G.}\ \bibnamefont {Nasibulin}}, \bibinfo {author}
  {\bibfnamefont {J.}~\bibnamefont {Becker}}, \bibinfo {author} {\bibfnamefont
  {J.}~\bibnamefont {Treu}}, \bibinfo {author} {\bibfnamefont {D.}~\bibnamefont
  {Ruhstorfer}}, \bibinfo {author} {\bibfnamefont {G.}~\bibnamefont
  {Koblm{\"u}ller}}, \bibinfo {author} {\bibfnamefont {E.~S.}\ \bibnamefont
  {Tikhonov}},\ and\ \bibinfo {author} {\bibfnamefont {V.~S.}\ \bibnamefont
  {Khrapai}},\ }\bibfield  {title} {\bibinfo {title} {Charge-neutral nonlocal
  response in superconductor-{{InAs}} nanowire hybrid devices},\ }\href
  {https://doi.org/10.1088/1361-6641/ac187b} {\bibfield  {journal} {\bibinfo
  {journal} {Semicond. Sci. Technol.}\ }\textbf {\bibinfo {volume} {36}},\
  \bibinfo {pages} {09LT04} (\bibinfo {year} {2021})}\BibitemShut {NoStop}%
\bibitem [{\citenamefont {Puglia}\ \emph {et~al.}(2021)\citenamefont {Puglia},
  \citenamefont {Martinez}, \citenamefont {M{\'e}nard}, \citenamefont
  {P{\"o}schl}, \citenamefont {Gronin}, \citenamefont {Gardner}, \citenamefont
  {Kallaher}, \citenamefont {Manfra}, \citenamefont {Marcus}, \citenamefont
  {Higginbotham},\ and\ \citenamefont {Casparis}}]{puglia2021}%
  \BibitemOpen
  \bibfield  {author} {\bibinfo {author} {\bibfnamefont {D.}~\bibnamefont
  {Puglia}}, \bibinfo {author} {\bibfnamefont {E.~A.}\ \bibnamefont
  {Martinez}}, \bibinfo {author} {\bibfnamefont {G.~C.}\ \bibnamefont
  {M{\'e}nard}}, \bibinfo {author} {\bibfnamefont {A.}~\bibnamefont
  {P{\"o}schl}}, \bibinfo {author} {\bibfnamefont {S.}~\bibnamefont {Gronin}},
  \bibinfo {author} {\bibfnamefont {G.~C.}\ \bibnamefont {Gardner}}, \bibinfo
  {author} {\bibfnamefont {R.}~\bibnamefont {Kallaher}}, \bibinfo {author}
  {\bibfnamefont {M.~J.}\ \bibnamefont {Manfra}}, \bibinfo {author}
  {\bibfnamefont {C.~M.}\ \bibnamefont {Marcus}}, \bibinfo {author}
  {\bibfnamefont {A.~P.}\ \bibnamefont {Higginbotham}},\ and\ \bibinfo {author}
  {\bibfnamefont {L.}~\bibnamefont {Casparis}},\ }\bibfield  {title} {\bibinfo
  {title} {Closing of the {{Induced Gap}} in a {{Hybrid
  Superconductor-Semiconductor Nanowire}}},\ }\href
  {https://doi.org/10.1103/PhysRevB.103.235201} {\bibfield  {journal} {\bibinfo
   {journal} {Physical Review B}\ }\textbf {\bibinfo {volume} {103}},\ \bibinfo
  {pages} {235201} (\bibinfo {year} {2021})}\BibitemShut {NoStop}%
\bibitem [{\citenamefont {Pan}\ \emph {et~al.}(2021)\citenamefont {Pan},
  \citenamefont {Sau},\ and\ \citenamefont {Das~Sarma}}]{pan2021}%
  \BibitemOpen
  \bibfield  {author} {\bibinfo {author} {\bibfnamefont {H.}~\bibnamefont
  {Pan}}, \bibinfo {author} {\bibfnamefont {J.~D.}\ \bibnamefont {Sau}},\ and\
  \bibinfo {author} {\bibfnamefont {S.}~\bibnamefont {Das~Sarma}},\ }\bibfield
  {title} {\bibinfo {title} {Three-terminal nonlocal conductance in
  {{Majorana}} nanowires: {{Distinguishing}} topological and trivial in
  realistic systems with disorder and inhomogeneous potential},\ }\href
  {https://doi.org/10.1103/PhysRevB.103.014513} {\bibfield  {journal} {\bibinfo
   {journal} {Physical Review B}\ }\textbf {\bibinfo {volume} {103}},\ \bibinfo
  {pages} {014513} (\bibinfo {year} {2021})}\BibitemShut {NoStop}%
\bibitem [{\citenamefont {Denisov}\ \emph {et~al.}(2022)\citenamefont
  {Denisov}, \citenamefont {Bubis}, \citenamefont {Piatrusha}, \citenamefont
  {Titova}, \citenamefont {Nasibulin}, \citenamefont {Becker}, \citenamefont
  {Treu}, \citenamefont {Ruhstorfer}, \citenamefont {Koblm{\"u}ller},
  \citenamefont {Tikhonov},\ and\ \citenamefont {Khrapai}}]{denisov2022}%
  \BibitemOpen
  \bibfield  {author} {\bibinfo {author} {\bibfnamefont {A.}~\bibnamefont
  {Denisov}}, \bibinfo {author} {\bibfnamefont {A.}~\bibnamefont {Bubis}},
  \bibinfo {author} {\bibfnamefont {S.}~\bibnamefont {Piatrusha}}, \bibinfo
  {author} {\bibfnamefont {N.}~\bibnamefont {Titova}}, \bibinfo {author}
  {\bibfnamefont {A.}~\bibnamefont {Nasibulin}}, \bibinfo {author}
  {\bibfnamefont {J.}~\bibnamefont {Becker}}, \bibinfo {author} {\bibfnamefont
  {J.}~\bibnamefont {Treu}}, \bibinfo {author} {\bibfnamefont {D.}~\bibnamefont
  {Ruhstorfer}}, \bibinfo {author} {\bibfnamefont {G.}~\bibnamefont
  {Koblm{\"u}ller}}, \bibinfo {author} {\bibfnamefont {E.}~\bibnamefont
  {Tikhonov}},\ and\ \bibinfo {author} {\bibfnamefont {V.}~\bibnamefont
  {Khrapai}},\ }\bibfield  {title} {\bibinfo {title} {Heat-{{Mode Excitation}}
  in a {{Proximity Superconductor}}},\ }\href
  {https://doi.org/10.3390/nano12091461} {\bibfield  {journal} {\bibinfo
  {journal} {Nanomaterials}\ }\textbf {\bibinfo {volume} {12}},\ \bibinfo
  {pages} {1461} (\bibinfo {year} {2022})}\BibitemShut {NoStop}%
\bibitem [{\citenamefont {Wang}\ \emph
  {et~al.}(2022{\natexlab{a}})\citenamefont {Wang}, \citenamefont {Dvir},
  \citenamefont {Van~Loo}, \citenamefont {Mazur}, \citenamefont {Gazibegovic},
  \citenamefont {Badawy}, \citenamefont {Bakkers}, \citenamefont
  {Kouwenhoven},\ and\ \citenamefont {De~Lange}}]{wang2022a}%
  \BibitemOpen
  \bibfield  {author} {\bibinfo {author} {\bibfnamefont {G.}~\bibnamefont
  {Wang}}, \bibinfo {author} {\bibfnamefont {T.}~\bibnamefont {Dvir}}, \bibinfo
  {author} {\bibfnamefont {N.}~\bibnamefont {Van~Loo}}, \bibinfo {author}
  {\bibfnamefont {G.~P.}\ \bibnamefont {Mazur}}, \bibinfo {author}
  {\bibfnamefont {S.}~\bibnamefont {Gazibegovic}}, \bibinfo {author}
  {\bibfnamefont {G.}~\bibnamefont {Badawy}}, \bibinfo {author} {\bibfnamefont
  {E.~P. A.~M.}\ \bibnamefont {Bakkers}}, \bibinfo {author} {\bibfnamefont
  {L.~P.}\ \bibnamefont {Kouwenhoven}},\ and\ \bibinfo {author} {\bibfnamefont
  {G.}~\bibnamefont {De~Lange}},\ }\bibfield  {title} {\bibinfo {title}
  {Nonlocal measurement of quasiparticle charge and energy relaxation in
  proximitized semiconductor nanowires using quantum dots},\ }\href
  {https://doi.org/10.1103/PhysRevB.106.064503} {\bibfield  {journal} {\bibinfo
   {journal} {Physical Review B}\ }\textbf {\bibinfo {volume} {106}},\ \bibinfo
  {pages} {064503} (\bibinfo {year} {2022}{\natexlab{a}})}\BibitemShut
  {NoStop}%
\bibitem [{\citenamefont {Kejriwal}\ and\ \citenamefont
  {Muralidharan}(2022)}]{kejriwal2022}%
  \BibitemOpen
  \bibfield  {author} {\bibinfo {author} {\bibfnamefont {A.}~\bibnamefont
  {Kejriwal}}\ and\ \bibinfo {author} {\bibfnamefont {B.}~\bibnamefont
  {Muralidharan}},\ }\bibfield  {title} {\bibinfo {title} {Nonlocal conductance
  and the detection of {{Majorana}} zero modes: {{Insights}} from von
  {{Neumann}} entropy},\ }\href {https://doi.org/10.1103/PhysRevB.105.L161403}
  {\bibfield  {journal} {\bibinfo  {journal} {Physical Review B}\ }\textbf
  {\bibinfo {volume} {105}},\ \bibinfo {pages} {L161403} (\bibinfo {year}
  {2022})}\BibitemShut {NoStop}%
\bibitem [{\citenamefont {Akhmerov}\ \emph {et~al.}(2011)\citenamefont
  {Akhmerov}, \citenamefont {Dahlhaus}, \citenamefont {Hassler}, \citenamefont
  {Wimmer},\ and\ \citenamefont {Beenakker}}]{akhmerov2011}%
  \BibitemOpen
  \bibfield  {author} {\bibinfo {author} {\bibfnamefont {A.~R.}\ \bibnamefont
  {Akhmerov}}, \bibinfo {author} {\bibfnamefont {J.~P.}\ \bibnamefont
  {Dahlhaus}}, \bibinfo {author} {\bibfnamefont {F.}~\bibnamefont {Hassler}},
  \bibinfo {author} {\bibfnamefont {M.}~\bibnamefont {Wimmer}},\ and\ \bibinfo
  {author} {\bibfnamefont {C.~W.~J.}\ \bibnamefont {Beenakker}},\ }\bibfield
  {title} {\bibinfo {title} {Quantized {{Conductance}} at the {{Majorana Phase
  Transition}} in a {{Disordered Superconducting Wire}}},\ }\href
  {https://doi.org/10.1103/PhysRevLett.106.057001} {\bibfield  {journal}
  {\bibinfo  {journal} {Physical Review Letters}\ }\textbf {\bibinfo {volume}
  {106}},\ \bibinfo {pages} {057001} (\bibinfo {year} {2011})}\BibitemShut
  {NoStop}%
\bibitem [{\citenamefont {Nishio}\ \emph {et~al.}(2011)\citenamefont {Nishio},
  \citenamefont {Kozakai}, \citenamefont {Amaha}, \citenamefont {Larsson},
  \citenamefont {Nilsson}, \citenamefont {Xu}, \citenamefont {Zhang},
  \citenamefont {Tateno}, \citenamefont {Takayanagi},\ and\ \citenamefont
  {Ishibashi}}]{nishio2011}%
  \BibitemOpen
  \bibfield  {author} {\bibinfo {author} {\bibfnamefont {T.}~\bibnamefont
  {Nishio}}, \bibinfo {author} {\bibfnamefont {T.}~\bibnamefont {Kozakai}},
  \bibinfo {author} {\bibfnamefont {S.}~\bibnamefont {Amaha}}, \bibinfo
  {author} {\bibfnamefont {M.}~\bibnamefont {Larsson}}, \bibinfo {author}
  {\bibfnamefont {H.~A.}\ \bibnamefont {Nilsson}}, \bibinfo {author}
  {\bibfnamefont {H.~Q.}\ \bibnamefont {Xu}}, \bibinfo {author} {\bibfnamefont
  {G.}~\bibnamefont {Zhang}}, \bibinfo {author} {\bibfnamefont
  {K.}~\bibnamefont {Tateno}}, \bibinfo {author} {\bibfnamefont
  {H.}~\bibnamefont {Takayanagi}},\ and\ \bibinfo {author} {\bibfnamefont
  {K.}~\bibnamefont {Ishibashi}},\ }\bibfield  {title} {\bibinfo {title}
  {Supercurrent through {{InAs}} nanowires with highly transparent
  superconducting contacts},\ }\href
  {https://doi.org/10.1088/0957-4484/22/44/445701} {\bibfield  {journal}
  {\bibinfo  {journal} {Nanotechnology}\ }\textbf {\bibinfo {volume} {22}},\
  \bibinfo {pages} {445701} (\bibinfo {year} {2011})}\BibitemShut {NoStop}%
\bibitem [{\citenamefont {Abay}\ \emph {et~al.}(2012)\citenamefont {Abay},
  \citenamefont {Nilsson}, \citenamefont {Wu}, \citenamefont {Xu},
  \citenamefont {Wilson},\ and\ \citenamefont {Delsing}}]{abay2012}%
  \BibitemOpen
  \bibfield  {author} {\bibinfo {author} {\bibfnamefont {S.}~\bibnamefont
  {Abay}}, \bibinfo {author} {\bibfnamefont {H.}~\bibnamefont {Nilsson}},
  \bibinfo {author} {\bibfnamefont {F.}~\bibnamefont {Wu}}, \bibinfo {author}
  {\bibfnamefont {H.}~\bibnamefont {Xu}}, \bibinfo {author} {\bibfnamefont
  {C.}~\bibnamefont {Wilson}},\ and\ \bibinfo {author} {\bibfnamefont
  {P.}~\bibnamefont {Delsing}},\ }\bibfield  {title} {\bibinfo {title} {High
  {{Critical-Current Superconductor-InAs Nanowire-Superconductor Junctions}}},\
  }\href {https://doi.org/10.1021/nl302740f} {\bibfield  {journal} {\bibinfo
  {journal} {Nano Letters}\ }\textbf {\bibinfo {volume} {12}},\ \bibinfo
  {pages} {5622} (\bibinfo {year} {2012})}\BibitemShut {NoStop}%
\bibitem [{\citenamefont {Abay}\ \emph {et~al.}(2014)\citenamefont {Abay},
  \citenamefont {Persson}, \citenamefont {Nilsson}, \citenamefont {Wu},
  \citenamefont {Xu}, \citenamefont {Fogelstr{\"o}m}, \citenamefont
  {Shumeiko},\ and\ \citenamefont {Delsing}}]{abay2014}%
  \BibitemOpen
  \bibfield  {author} {\bibinfo {author} {\bibfnamefont {S.}~\bibnamefont
  {Abay}}, \bibinfo {author} {\bibfnamefont {D.}~\bibnamefont {Persson}},
  \bibinfo {author} {\bibfnamefont {H.}~\bibnamefont {Nilsson}}, \bibinfo
  {author} {\bibfnamefont {F.}~\bibnamefont {Wu}}, \bibinfo {author}
  {\bibfnamefont {H.~Q.}\ \bibnamefont {Xu}}, \bibinfo {author} {\bibfnamefont
  {M.}~\bibnamefont {Fogelstr{\"o}m}}, \bibinfo {author} {\bibfnamefont
  {V.}~\bibnamefont {Shumeiko}},\ and\ \bibinfo {author} {\bibfnamefont
  {P.}~\bibnamefont {Delsing}},\ }\bibfield  {title} {\bibinfo {title} {Charge
  transport in {{InAs}} nanowire {{Josephson}} junctions},\ }\href
  {https://doi.org/10.1103/PhysRevB.89.214508} {\bibfield  {journal} {\bibinfo
  {journal} {Physical Review B}\ }\textbf {\bibinfo {volume} {89}},\ \bibinfo
  {pages} {214508} (\bibinfo {year} {2014})}\BibitemShut {NoStop}%
\bibitem [{\citenamefont {Paajaste}\ \emph {et~al.}(2015)\citenamefont
  {Paajaste}, \citenamefont {Amado}, \citenamefont {Roddaro}, \citenamefont
  {Bergeret}, \citenamefont {Ercolani}, \citenamefont {Sorba},\ and\
  \citenamefont {Giazotto}}]{paajaste2015}%
  \BibitemOpen
  \bibfield  {author} {\bibinfo {author} {\bibfnamefont {J.}~\bibnamefont
  {Paajaste}}, \bibinfo {author} {\bibfnamefont {M.}~\bibnamefont {Amado}},
  \bibinfo {author} {\bibfnamefont {S.}~\bibnamefont {Roddaro}}, \bibinfo
  {author} {\bibfnamefont {F.~S.}\ \bibnamefont {Bergeret}}, \bibinfo {author}
  {\bibfnamefont {D.}~\bibnamefont {Ercolani}}, \bibinfo {author}
  {\bibfnamefont {L.}~\bibnamefont {Sorba}},\ and\ \bibinfo {author}
  {\bibfnamefont {F.}~\bibnamefont {Giazotto}},\ }\bibfield  {title} {\bibinfo
  {title} {Pb/{{InAs Nanowire Josephson Junction}} with {{High Critical
  Current}} and {{Magnetic Flux Focusing}}},\ }\href
  {https://doi.org/10.1021/nl504544s} {\bibfield  {journal} {\bibinfo
  {journal} {Nano Letters}\ }\textbf {\bibinfo {volume} {15}},\ \bibinfo
  {pages} {1803} (\bibinfo {year} {2015})}\BibitemShut {NoStop}%
\bibitem [{\citenamefont {Perla}(2021)}]{perla2021}%
  \BibitemOpen
  \bibfield  {author} {\bibinfo {author} {\bibfnamefont {P.}~\bibnamefont
  {Perla}},\ }\bibfield  {title} {\bibinfo {title} {Fully in situ
  {{Nb}}/{{InAs-nanowire Josephson}} junctions by selective-area growth and
  shadow evaporation},\ }\href {https://doi.org/10.1039/d0na00999g} {\bibfield
  {journal} {\bibinfo  {journal} {Nanoscale Advances}\ }\textbf {\bibinfo
  {volume} {3}},\ \bibinfo {pages} {1413} (\bibinfo {year} {2021})}\BibitemShut
  {NoStop}%
\bibitem [{\citenamefont {Kousar}\ \emph {et~al.}(2022)\citenamefont {Kousar},
  \citenamefont {Carrad}, \citenamefont {Stampfer}, \citenamefont {Krogstrup},
  \citenamefont {Nyg{\aa}rd},\ and\ \citenamefont {Jespersen}}]{kousar2022}%
  \BibitemOpen
  \bibfield  {author} {\bibinfo {author} {\bibfnamefont {B.}~\bibnamefont
  {Kousar}}, \bibinfo {author} {\bibfnamefont {D.~J.}\ \bibnamefont {Carrad}},
  \bibinfo {author} {\bibfnamefont {L.}~\bibnamefont {Stampfer}}, \bibinfo
  {author} {\bibfnamefont {P.}~\bibnamefont {Krogstrup}}, \bibinfo {author}
  {\bibfnamefont {J.}~\bibnamefont {Nyg{\aa}rd}},\ and\ \bibinfo {author}
  {\bibfnamefont {T.~S.}\ \bibnamefont {Jespersen}},\ }\bibfield  {title}
  {\bibinfo {title} {{{InAs}}/{{MoRe Hybrid Semiconductor}}/{{Superconductor
  Nanowire Devices}}},\ }\href {https://doi.org/10.1021/acs.nanolett.2c02532}
  {\bibfield  {journal} {\bibinfo  {journal} {Nano Letters}\ }\textbf {\bibinfo
  {volume} {22}},\ \bibinfo {pages} {8845} (\bibinfo {year}
  {2022})}\BibitemShut {NoStop}%
\bibitem [{\citenamefont {Spanton}\ \emph {et~al.}(2017)\citenamefont
  {Spanton}, \citenamefont {Deng}, \citenamefont {Vaitiek{\.e}nas},
  \citenamefont {Krogstrup}, \citenamefont {Nyg{\aa}rd}, \citenamefont
  {Marcus},\ and\ \citenamefont {Moler}}]{spanton2017}%
  \BibitemOpen
  \bibfield  {author} {\bibinfo {author} {\bibfnamefont {E.~M.}\ \bibnamefont
  {Spanton}}, \bibinfo {author} {\bibfnamefont {M.}~\bibnamefont {Deng}},
  \bibinfo {author} {\bibfnamefont {S.}~\bibnamefont {Vaitiek{\.e}nas}},
  \bibinfo {author} {\bibfnamefont {P.}~\bibnamefont {Krogstrup}}, \bibinfo
  {author} {\bibfnamefont {J.}~\bibnamefont {Nyg{\aa}rd}}, \bibinfo {author}
  {\bibfnamefont {C.~M.}\ \bibnamefont {Marcus}},\ and\ \bibinfo {author}
  {\bibfnamefont {K.~A.}\ \bibnamefont {Moler}},\ }\bibfield  {title} {\bibinfo
  {title} {Current{\textendash}phase relations of few-mode {{InAs}} nanowire
  {{Josephson}} junctions},\ }\href {https://doi.org/10.1038/nphys4224}
  {\bibfield  {journal} {\bibinfo  {journal} {Nature Physics}\ }\textbf
  {\bibinfo {volume} {13}},\ \bibinfo {pages} {1177} (\bibinfo {year}
  {2017})}\BibitemShut {NoStop}%
\bibitem [{\citenamefont {Hart}(2019)}]{hart2019}%
  \BibitemOpen
  \bibfield  {author} {\bibinfo {author} {\bibfnamefont {S.}~\bibnamefont
  {Hart}},\ }\bibfield  {title} {\bibinfo {title} {Current-phase relations of
  {{InAs}} nanowire {{Josephson}} junctions: {{From}} interacting to multimode
  regimes},\ }\href {https://doi.org/10.1103/PhysRevB.100.064523} {\bibfield
  {journal} {\bibinfo  {journal} {Physical Review B}\ }\textbf {\bibinfo
  {volume} {100}},\ \bibinfo {pages} {064523} (\bibinfo {year}
  {2019})}\BibitemShut {NoStop}%
\bibitem [{\citenamefont {Das}\ \emph {et~al.}(2012{\natexlab{a}})\citenamefont
  {Das}, \citenamefont {Ronen}, \citenamefont {Most}, \citenamefont {Oreg},
  \citenamefont {Heiblum},\ and\ \citenamefont {Shtrikman}}]{das2012}%
  \BibitemOpen
  \bibfield  {author} {\bibinfo {author} {\bibfnamefont {A.}~\bibnamefont
  {Das}}, \bibinfo {author} {\bibfnamefont {Y.}~\bibnamefont {Ronen}}, \bibinfo
  {author} {\bibfnamefont {Y.}~\bibnamefont {Most}}, \bibinfo {author}
  {\bibfnamefont {Y.}~\bibnamefont {Oreg}}, \bibinfo {author} {\bibfnamefont
  {M.}~\bibnamefont {Heiblum}},\ and\ \bibinfo {author} {\bibfnamefont
  {H.}~\bibnamefont {Shtrikman}},\ }\bibfield  {title} {\bibinfo {title}
  {Zero-bias peaks and splitting in an {{Al}}{\textendash}{{InAs}} nanowire
  topological superconductor as a signature of {{Majorana}} fermions},\ }\href
  {https://doi.org/10.1038/nphys2479} {\bibfield  {journal} {\bibinfo
  {journal} {Nature Physics}\ }\textbf {\bibinfo {volume} {8}},\ \bibinfo
  {pages} {887} (\bibinfo {year} {2012}{\natexlab{a}})}\BibitemShut {NoStop}%
\bibitem [{\citenamefont {Mourik}\ \emph {et~al.}(2012)\citenamefont {Mourik},
  \citenamefont {Zuo}, \citenamefont {Frolov}, \citenamefont {Plissard},
  \citenamefont {Bakkers},\ and\ \citenamefont {Kouwenhoven}}]{mourik2012}%
  \BibitemOpen
  \bibfield  {author} {\bibinfo {author} {\bibfnamefont {V.}~\bibnamefont
  {Mourik}}, \bibinfo {author} {\bibfnamefont {K.}~\bibnamefont {Zuo}},
  \bibinfo {author} {\bibfnamefont {S.~M.}\ \bibnamefont {Frolov}}, \bibinfo
  {author} {\bibfnamefont {S.~R.}\ \bibnamefont {Plissard}}, \bibinfo {author}
  {\bibfnamefont {E.~P. A.~M.}\ \bibnamefont {Bakkers}},\ and\ \bibinfo
  {author} {\bibfnamefont {L.~P.}\ \bibnamefont {Kouwenhoven}},\ }\bibfield
  {title} {\bibinfo {title} {Signatures of {{Majorana Fermions}} in {{Hybrid
  Superconductor-Semiconductor Nanowire Devices}}},\ }\href
  {https://doi.org/10.1126/science.1222360} {\bibfield  {journal} {\bibinfo
  {journal} {Science}\ }\textbf {\bibinfo {volume} {336}},\ \bibinfo {pages}
  {1003} (\bibinfo {year} {2012})}\BibitemShut {NoStop}%
\bibitem [{\citenamefont {Vaitiek{\.e}nas}\ \emph
  {et~al.}(2020{\natexlab{a}})\citenamefont {Vaitiek{\.e}nas}, \citenamefont
  {Winkler}, \citenamefont {Van~Heck}, \citenamefont {Karzig}, \citenamefont
  {Deng}, \citenamefont {Flensberg}, \citenamefont {Glazman}, \citenamefont
  {Nayak}, \citenamefont {Krogstrup}, \citenamefont {Lutchyn},\ and\
  \citenamefont {Marcus}}]{vaitiekenas2020a}%
  \BibitemOpen
  \bibfield  {author} {\bibinfo {author} {\bibfnamefont {S.}~\bibnamefont
  {Vaitiek{\.e}nas}}, \bibinfo {author} {\bibfnamefont {G.~W.}\ \bibnamefont
  {Winkler}}, \bibinfo {author} {\bibfnamefont {B.}~\bibnamefont {Van~Heck}},
  \bibinfo {author} {\bibfnamefont {T.}~\bibnamefont {Karzig}}, \bibinfo
  {author} {\bibfnamefont {M.-T.}\ \bibnamefont {Deng}}, \bibinfo {author}
  {\bibfnamefont {K.}~\bibnamefont {Flensberg}}, \bibinfo {author}
  {\bibfnamefont {L.~I.}\ \bibnamefont {Glazman}}, \bibinfo {author}
  {\bibfnamefont {C.}~\bibnamefont {Nayak}}, \bibinfo {author} {\bibfnamefont
  {P.}~\bibnamefont {Krogstrup}}, \bibinfo {author} {\bibfnamefont {R.~M.}\
  \bibnamefont {Lutchyn}},\ and\ \bibinfo {author} {\bibfnamefont {C.~M.}\
  \bibnamefont {Marcus}},\ }\bibfield  {title} {\bibinfo {title} {Flux-induced
  topological superconductivity in full-shell nanowires},\ }\href
  {https://doi.org/10.1126/science.aav3392} {\bibfield  {journal} {\bibinfo
  {journal} {Science}\ }\textbf {\bibinfo {volume} {367}},\ \bibinfo {pages}
  {eaav3392} (\bibinfo {year} {2020}{\natexlab{a}})}\BibitemShut {NoStop}%
\bibitem [{\citenamefont {Vaitiek{\.e}nas}\ \emph {et~al.}(2021)\citenamefont
  {Vaitiek{\.e}nas}, \citenamefont {Liu}, \citenamefont {Krogstrup},\ and\
  \citenamefont {Marcus}}]{vaitiekenas2021}%
  \BibitemOpen
  \bibfield  {author} {\bibinfo {author} {\bibfnamefont {S.}~\bibnamefont
  {Vaitiek{\.e}nas}}, \bibinfo {author} {\bibfnamefont {Y.}~\bibnamefont
  {Liu}}, \bibinfo {author} {\bibfnamefont {P.}~\bibnamefont {Krogstrup}},\
  and\ \bibinfo {author} {\bibfnamefont {C.~M.}\ \bibnamefont {Marcus}},\
  }\bibfield  {title} {\bibinfo {title} {Zero-bias peaks at zero magnetic field
  in ferromagnetic hybrid nanowires},\ }\href
  {https://doi.org/10.1038/s41567-020-1017-3} {\bibfield  {journal} {\bibinfo
  {journal} {Nature Physics}\ }\textbf {\bibinfo {volume} {17}},\ \bibinfo
  {pages} {43} (\bibinfo {year} {2021})}\BibitemShut {NoStop}%
\bibitem [{\citenamefont {Valentini}\ \emph {et~al.}(2021)\citenamefont
  {Valentini}, \citenamefont {Pe{\~n}aranda}, \citenamefont {Hofmann},
  \citenamefont {Brauns}, \citenamefont {Hauschild}, \citenamefont {Krogstrup},
  \citenamefont {{San-Jose}}, \citenamefont {Prada}, \citenamefont {Aguado},\
  and\ \citenamefont {Katsaros}}]{valentini2021}%
  \BibitemOpen
  \bibfield  {author} {\bibinfo {author} {\bibfnamefont {M.}~\bibnamefont
  {Valentini}}, \bibinfo {author} {\bibfnamefont {F.}~\bibnamefont
  {Pe{\~n}aranda}}, \bibinfo {author} {\bibfnamefont {A.}~\bibnamefont
  {Hofmann}}, \bibinfo {author} {\bibfnamefont {M.}~\bibnamefont {Brauns}},
  \bibinfo {author} {\bibfnamefont {R.}~\bibnamefont {Hauschild}}, \bibinfo
  {author} {\bibfnamefont {P.}~\bibnamefont {Krogstrup}}, \bibinfo {author}
  {\bibfnamefont {P.}~\bibnamefont {{San-Jose}}}, \bibinfo {author}
  {\bibfnamefont {E.}~\bibnamefont {Prada}}, \bibinfo {author} {\bibfnamefont
  {R.}~\bibnamefont {Aguado}},\ and\ \bibinfo {author} {\bibfnamefont
  {G.}~\bibnamefont {Katsaros}},\ }\bibfield  {title} {\bibinfo {title}
  {Nontopological zero-bias peaks in full-shell nanowires induced by
  flux-tunable {{Andreev}} states},\ }\href
  {https://doi.org/10.1126/science.abf1513} {\bibfield  {journal} {\bibinfo
  {journal} {Science}\ }\textbf {\bibinfo {volume} {373}},\ \bibinfo {pages}
  {82} (\bibinfo {year} {2021})}\BibitemShut {NoStop}%
\bibitem [{\citenamefont {Hofstetter}\ \emph {et~al.}(2009)\citenamefont
  {Hofstetter}, \citenamefont {Csonka}, \citenamefont {Nyg{\aa}rd},\ and\
  \citenamefont {Sch{\"o}nenberger}}]{hofstetter2009}%
  \BibitemOpen
  \bibfield  {author} {\bibinfo {author} {\bibfnamefont {L.}~\bibnamefont
  {Hofstetter}}, \bibinfo {author} {\bibfnamefont {S.}~\bibnamefont {Csonka}},
  \bibinfo {author} {\bibfnamefont {J.}~\bibnamefont {Nyg{\aa}rd}},\ and\
  \bibinfo {author} {\bibfnamefont {C.}~\bibnamefont {Sch{\"o}nenberger}},\
  }\bibfield  {title} {\bibinfo {title} {Cooper pair splitter realized in a
  two-quantum-dot {{Y-junction}}},\ }\href
  {https://doi.org/10.1038/nature08432} {\bibfield  {journal} {\bibinfo
  {journal} {Nature}\ }\textbf {\bibinfo {volume} {461}},\ \bibinfo {pages}
  {960} (\bibinfo {year} {2009})}\BibitemShut {NoStop}%
\bibitem [{\citenamefont {Herrmann}\ \emph {et~al.}(2010)\citenamefont
  {Herrmann}, \citenamefont {Portier}, \citenamefont {Roche}, \citenamefont
  {Yeyati}, \citenamefont {Kontos},\ and\ \citenamefont
  {Strunk}}]{herrmann2010}%
  \BibitemOpen
  \bibfield  {author} {\bibinfo {author} {\bibfnamefont {L.~G.}\ \bibnamefont
  {Herrmann}}, \bibinfo {author} {\bibfnamefont {F.}~\bibnamefont {Portier}},
  \bibinfo {author} {\bibfnamefont {P.}~\bibnamefont {Roche}}, \bibinfo
  {author} {\bibfnamefont {A.~L.}\ \bibnamefont {Yeyati}}, \bibinfo {author}
  {\bibfnamefont {T.}~\bibnamefont {Kontos}},\ and\ \bibinfo {author}
  {\bibfnamefont {C.}~\bibnamefont {Strunk}},\ }\bibfield  {title} {\bibinfo
  {title} {Carbon {{Nanotubes}} as {{Cooper-Pair Beam Splitters}}},\ }\href
  {https://doi.org/10.1103/PhysRevLett.104.026801} {\bibfield  {journal}
  {\bibinfo  {journal} {Physical Review Letters}\ }\textbf {\bibinfo {volume}
  {104}},\ \bibinfo {pages} {026801} (\bibinfo {year} {2010})}\BibitemShut
  {NoStop}%
\bibitem [{\citenamefont {Das}\ \emph {et~al.}(2012{\natexlab{b}})\citenamefont
  {Das}, \citenamefont {Ronen}, \citenamefont {Heiblum}, \citenamefont
  {Mahalu}, \citenamefont {Kretinin},\ and\ \citenamefont
  {Shtrikman}}]{das2012a}%
  \BibitemOpen
  \bibfield  {author} {\bibinfo {author} {\bibfnamefont {A.}~\bibnamefont
  {Das}}, \bibinfo {author} {\bibfnamefont {Y.}~\bibnamefont {Ronen}}, \bibinfo
  {author} {\bibfnamefont {M.}~\bibnamefont {Heiblum}}, \bibinfo {author}
  {\bibfnamefont {D.}~\bibnamefont {Mahalu}}, \bibinfo {author} {\bibfnamefont
  {A.~V.}\ \bibnamefont {Kretinin}},\ and\ \bibinfo {author} {\bibfnamefont
  {H.}~\bibnamefont {Shtrikman}},\ }\bibfield  {title} {\bibinfo {title}
  {High-efficiency {{Cooper}} pair splitting demonstrated by two-particle
  conductance resonance and positive noise cross-correlation},\ }\href
  {https://doi.org/10.1038/ncomms2169} {\bibfield  {journal} {\bibinfo
  {journal} {Nature Communications}\ }\textbf {\bibinfo {volume} {3}},\
  \bibinfo {pages} {1165} (\bibinfo {year} {2012}{\natexlab{b}})}\BibitemShut
  {NoStop}%
\bibitem [{\citenamefont {Baba}\ \emph {et~al.}(2018)\citenamefont {Baba},
  \citenamefont {J{\"u}nger}, \citenamefont {Matsuo}, \citenamefont
  {Baumgartner}, \citenamefont {Sato}, \citenamefont {Kamata}, \citenamefont
  {Li}, \citenamefont {Jeppesen}, \citenamefont {Samuelson}, \citenamefont
  {Xu}, \citenamefont {Sch{\"o}nenberger},\ and\ \citenamefont
  {Tarucha}}]{baba2018}%
  \BibitemOpen
  \bibfield  {author} {\bibinfo {author} {\bibfnamefont {S.}~\bibnamefont
  {Baba}}, \bibinfo {author} {\bibfnamefont {C.}~\bibnamefont {J{\"u}nger}},
  \bibinfo {author} {\bibfnamefont {S.}~\bibnamefont {Matsuo}}, \bibinfo
  {author} {\bibfnamefont {A.}~\bibnamefont {Baumgartner}}, \bibinfo {author}
  {\bibfnamefont {Y.}~\bibnamefont {Sato}}, \bibinfo {author} {\bibfnamefont
  {H.}~\bibnamefont {Kamata}}, \bibinfo {author} {\bibfnamefont
  {K.}~\bibnamefont {Li}}, \bibinfo {author} {\bibfnamefont {S.}~\bibnamefont
  {Jeppesen}}, \bibinfo {author} {\bibfnamefont {L.}~\bibnamefont {Samuelson}},
  \bibinfo {author} {\bibfnamefont {H.}~\bibnamefont {Xu}}, \bibinfo {author}
  {\bibfnamefont {C.}~\bibnamefont {Sch{\"o}nenberger}},\ and\ \bibinfo
  {author} {\bibfnamefont {S.}~\bibnamefont {Tarucha}},\ }\bibfield  {title}
  {\bibinfo {title} {Cooper-pair splitting in two parallel {{InAs}}
  nanowires},\ }\href {https://doi.org/10.1088/1367-2630/aac74e} {\bibfield
  {journal} {\bibinfo  {journal} {New Journal of Physics}\ }\textbf {\bibinfo
  {volume} {20}},\ \bibinfo {pages} {063021} (\bibinfo {year}
  {2018})}\BibitemShut {NoStop}%
\bibitem [{\citenamefont {Wang}\ \emph
  {et~al.}(2022{\natexlab{b}})\citenamefont {Wang}, \citenamefont {Dvir},
  \citenamefont {Mazur}, \citenamefont {Liu}, \citenamefont {Van~Loo},
  \citenamefont {Ten~Haaf}, \citenamefont {Bordin}, \citenamefont
  {Gazibegovic}, \citenamefont {Badawy}, \citenamefont {Bakkers}, \citenamefont
  {Wimmer},\ and\ \citenamefont {Kouwenhoven}}]{wang2022}%
  \BibitemOpen
  \bibfield  {author} {\bibinfo {author} {\bibfnamefont {G.}~\bibnamefont
  {Wang}}, \bibinfo {author} {\bibfnamefont {T.}~\bibnamefont {Dvir}}, \bibinfo
  {author} {\bibfnamefont {G.~P.}\ \bibnamefont {Mazur}}, \bibinfo {author}
  {\bibfnamefont {C.-X.}\ \bibnamefont {Liu}}, \bibinfo {author} {\bibfnamefont
  {N.}~\bibnamefont {Van~Loo}}, \bibinfo {author} {\bibfnamefont {S.~L.~D.}\
  \bibnamefont {Ten~Haaf}}, \bibinfo {author} {\bibfnamefont {A.}~\bibnamefont
  {Bordin}}, \bibinfo {author} {\bibfnamefont {S.}~\bibnamefont {Gazibegovic}},
  \bibinfo {author} {\bibfnamefont {G.}~\bibnamefont {Badawy}}, \bibinfo
  {author} {\bibfnamefont {E.~P. A.~M.}\ \bibnamefont {Bakkers}}, \bibinfo
  {author} {\bibfnamefont {M.}~\bibnamefont {Wimmer}},\ and\ \bibinfo {author}
  {\bibfnamefont {L.~P.}\ \bibnamefont {Kouwenhoven}},\ }\bibfield  {title}
  {\bibinfo {title} {Singlet and triplet {{Cooper}} pair splitting in hybrid
  superconducting nanowires},\ }\href
  {https://doi.org/10.1038/s41586-022-05352-2} {\bibfield  {journal} {\bibinfo
  {journal} {Nature}\ }\textbf {\bibinfo {volume} {612}},\ \bibinfo {pages}
  {448} (\bibinfo {year} {2022}{\natexlab{b}})}\BibitemShut {NoStop}%
\bibitem [{\citenamefont {Bordoloi}\ \emph {et~al.}(2022)\citenamefont
  {Bordoloi}, \citenamefont {Zannier}, \citenamefont {Sorba}, \citenamefont
  {Sch{\"o}nenberger},\ and\ \citenamefont {Baumgartner}}]{bordoloi2022}%
  \BibitemOpen
  \bibfield  {author} {\bibinfo {author} {\bibfnamefont {A.}~\bibnamefont
  {Bordoloi}}, \bibinfo {author} {\bibfnamefont {V.}~\bibnamefont {Zannier}},
  \bibinfo {author} {\bibfnamefont {L.}~\bibnamefont {Sorba}}, \bibinfo
  {author} {\bibfnamefont {C.}~\bibnamefont {Sch{\"o}nenberger}},\ and\
  \bibinfo {author} {\bibfnamefont {A.}~\bibnamefont {Baumgartner}},\
  }\bibfield  {title} {\bibinfo {title} {Spin cross-correlation experiments in
  an electron entangler},\ }\href {https://doi.org/10.1038/s41586-022-05436-z}
  {\bibfield  {journal} {\bibinfo  {journal} {Nature}\ }\textbf {\bibinfo
  {volume} {612}},\ \bibinfo {pages} {454} (\bibinfo {year}
  {2022})}\BibitemShut {NoStop}%
\bibitem [{\citenamefont {Scher{\"u}bl}\ \emph {et~al.}(2022)\citenamefont
  {Scher{\"u}bl}, \citenamefont {F{\"u}l{\"o}p}, \citenamefont {Gramich},
  \citenamefont {P{\'a}lyi}, \citenamefont {Sch{\"o}nenberger}, \citenamefont
  {Nyg{\aa}rd},\ and\ \citenamefont {Csonka}}]{scherubl2022}%
  \BibitemOpen
  \bibfield  {author} {\bibinfo {author} {\bibfnamefont {Z.}~\bibnamefont
  {Scher{\"u}bl}}, \bibinfo {author} {\bibfnamefont {G.}~\bibnamefont
  {F{\"u}l{\"o}p}}, \bibinfo {author} {\bibfnamefont {J.}~\bibnamefont
  {Gramich}}, \bibinfo {author} {\bibfnamefont {A.}~\bibnamefont {P{\'a}lyi}},
  \bibinfo {author} {\bibfnamefont {C.}~\bibnamefont {Sch{\"o}nenberger}},
  \bibinfo {author} {\bibfnamefont {J.}~\bibnamefont {Nyg{\aa}rd}},\ and\
  \bibinfo {author} {\bibfnamefont {S.}~\bibnamefont {Csonka}},\ }\bibfield
  {title} {\bibinfo {title} {From {{Cooper}} pair splitting to the non-local
  spectroscopy of a {{Shiba}} state},\ }\href
  {https://doi.org/10.1103/PhysRevResearch.4.023143} {\bibfield  {journal}
  {\bibinfo  {journal} {Physical Review Research}\ }\textbf {\bibinfo {volume}
  {4}},\ \bibinfo {pages} {023143} (\bibinfo {year} {2022})}\BibitemShut
  {NoStop}%
\bibitem [{\citenamefont {Vaitiek{\.e}nas}\ \emph
  {et~al.}(2020{\natexlab{b}})\citenamefont {Vaitiek{\.e}nas}, \citenamefont
  {Krogstrup},\ and\ \citenamefont {Marcus}}]{vaitiekenas2020}%
  \BibitemOpen
  \bibfield  {author} {\bibinfo {author} {\bibfnamefont {S.}~\bibnamefont
  {Vaitiek{\.e}nas}}, \bibinfo {author} {\bibfnamefont {P.}~\bibnamefont
  {Krogstrup}},\ and\ \bibinfo {author} {\bibfnamefont {C.~M.}\ \bibnamefont
  {Marcus}},\ }\bibfield  {title} {\bibinfo {title} {Anomalous metallic phase
  in tunable destructive superconductors},\ }\href
  {https://doi.org/10.1103/PhysRevB.101.060507} {\bibfield  {journal} {\bibinfo
   {journal} {Physical Review B}\ }\textbf {\bibinfo {volume} {101}},\ \bibinfo
  {pages} {060507} (\bibinfo {year} {2020}{\natexlab{b}})}\BibitemShut
  {NoStop}%
\bibitem [{\citenamefont {Keizer}\ \emph {et~al.}(2006)\citenamefont {Keizer},
  \citenamefont {Flokstra}, \citenamefont {Aarts},\ and\ \citenamefont
  {Klapwijk}}]{keizer2006}%
  \BibitemOpen
  \bibfield  {author} {\bibinfo {author} {\bibfnamefont {R.~S.}\ \bibnamefont
  {Keizer}}, \bibinfo {author} {\bibfnamefont {M.~G.}\ \bibnamefont
  {Flokstra}}, \bibinfo {author} {\bibfnamefont {J.}~\bibnamefont {Aarts}},\
  and\ \bibinfo {author} {\bibfnamefont {T.~M.}\ \bibnamefont {Klapwijk}},\
  }\bibfield  {title} {\bibinfo {title} {Critical {{Voltage}} of a {{Mesoscopic
  Superconductor}}},\ }\href {https://doi.org/10.1103/PhysRevLett.96.147002}
  {\bibfield  {journal} {\bibinfo  {journal} {Physical Review Letters}\
  }\textbf {\bibinfo {volume} {96}},\ \bibinfo {pages} {147002} (\bibinfo
  {year} {2006})}\BibitemShut {NoStop}%
\bibitem [{\citenamefont {Snyman}\ and\ \citenamefont
  {Nazarov}(2009)}]{snyman2009}%
  \BibitemOpen
  \bibfield  {author} {\bibinfo {author} {\bibfnamefont {I.}~\bibnamefont
  {Snyman}}\ and\ \bibinfo {author} {\bibfnamefont {{\relax Yu}.~V.}\
  \bibnamefont {Nazarov}},\ }\bibfield  {title} {\bibinfo {title} {Bistability
  in voltage-biased
  normal-metal/insulator/superconductor/insulator/normal-metal structures},\
  }\href {https://doi.org/10.1103/PhysRevB.79.014510} {\bibfield  {journal}
  {\bibinfo  {journal} {Physical Review B}\ }\textbf {\bibinfo {volume} {79}},\
  \bibinfo {pages} {014510} (\bibinfo {year} {2009})}\BibitemShut {NoStop}%
\bibitem [{\citenamefont {Huard}\ \emph {et~al.}(2007)\citenamefont {Huard},
  \citenamefont {Pothier}, \citenamefont {Esteve},\ and\ \citenamefont
  {Nagaev}}]{huard2007}%
  \BibitemOpen
  \bibfield  {author} {\bibinfo {author} {\bibfnamefont {B.}~\bibnamefont
  {Huard}}, \bibinfo {author} {\bibfnamefont {H.}~\bibnamefont {Pothier}},
  \bibinfo {author} {\bibfnamefont {D.}~\bibnamefont {Esteve}},\ and\ \bibinfo
  {author} {\bibfnamefont {K.~E.}\ \bibnamefont {Nagaev}},\ }\bibfield  {title}
  {\bibinfo {title} {Electron heating in metallic resistors at sub-{{Kelvin}}
  temperature},\ }\href {https://doi.org/10.1103/PhysRevB.76.165426} {\bibfield
   {journal} {\bibinfo  {journal} {Physical Review B}\ }\textbf {\bibinfo
  {volume} {76}},\ \bibinfo {pages} {165426} (\bibinfo {year}
  {2007})}\BibitemShut {NoStop}%
\bibitem [{\citenamefont {Sivre}\ \emph {et~al.}(2018)\citenamefont {Sivre},
  \citenamefont {Anthore}, \citenamefont {Parmentier}, \citenamefont {Cavanna},
  \citenamefont {Gennser}, \citenamefont {Ouerghi}, \citenamefont {Jin},\ and\
  \citenamefont {Pierre}}]{sivre2018}%
  \BibitemOpen
  \bibfield  {author} {\bibinfo {author} {\bibfnamefont {E.}~\bibnamefont
  {Sivre}}, \bibinfo {author} {\bibfnamefont {A.}~\bibnamefont {Anthore}},
  \bibinfo {author} {\bibfnamefont {F.~D.}\ \bibnamefont {Parmentier}},
  \bibinfo {author} {\bibfnamefont {A.}~\bibnamefont {Cavanna}}, \bibinfo
  {author} {\bibfnamefont {U.}~\bibnamefont {Gennser}}, \bibinfo {author}
  {\bibfnamefont {A.}~\bibnamefont {Ouerghi}}, \bibinfo {author} {\bibfnamefont
  {Y.}~\bibnamefont {Jin}},\ and\ \bibinfo {author} {\bibfnamefont
  {F.}~\bibnamefont {Pierre}},\ }\bibfield  {title} {\bibinfo {title} {Heat
  {{Coulomb}} blockade of one ballistic channel},\ }\href
  {https://doi.org/10.1038/nphys4280} {\bibfield  {journal} {\bibinfo
  {journal} {Nature Physics}\ }\textbf {\bibinfo {volume} {14}},\ \bibinfo
  {pages} {145} (\bibinfo {year} {2018})}\BibitemShut {NoStop}%
\bibitem [{\citenamefont {Rosenblatt}\ \emph {et~al.}(2020)\citenamefont
  {Rosenblatt}, \citenamefont {Konyzheva}, \citenamefont {Lafont},
  \citenamefont {Schiller}, \citenamefont {Park}, \citenamefont {Snizhko},
  \citenamefont {Heiblum}, \citenamefont {Oreg},\ and\ \citenamefont
  {Umansky}}]{rosenblatt2020}%
  \BibitemOpen
  \bibfield  {author} {\bibinfo {author} {\bibfnamefont {A.}~\bibnamefont
  {Rosenblatt}}, \bibinfo {author} {\bibfnamefont {S.}~\bibnamefont
  {Konyzheva}}, \bibinfo {author} {\bibfnamefont {F.}~\bibnamefont {Lafont}},
  \bibinfo {author} {\bibfnamefont {N.}~\bibnamefont {Schiller}}, \bibinfo
  {author} {\bibfnamefont {J.}~\bibnamefont {Park}}, \bibinfo {author}
  {\bibfnamefont {K.}~\bibnamefont {Snizhko}}, \bibinfo {author} {\bibfnamefont
  {M.}~\bibnamefont {Heiblum}}, \bibinfo {author} {\bibfnamefont
  {Y.}~\bibnamefont {Oreg}},\ and\ \bibinfo {author} {\bibfnamefont
  {V.}~\bibnamefont {Umansky}},\ }\bibfield  {title} {\bibinfo {title} {Energy
  {{Relaxation}} in {{Edge Modes}} in the {{Quantum Hall Effect}}},\ }\href
  {https://doi.org/10.1103/PhysRevLett.125.256803} {\bibfield  {journal}
  {\bibinfo  {journal} {Physical Review Letters}\ }\textbf {\bibinfo {volume}
  {125}},\ \bibinfo {pages} {256803} (\bibinfo {year} {2020})}\BibitemShut
  {NoStop}%
\bibitem [{\citenamefont {Vercruyssen}\ \emph {et~al.}(2012)\citenamefont
  {Vercruyssen}, \citenamefont {Verhagen}, \citenamefont {Flokstra},
  \citenamefont {Pekola},\ and\ \citenamefont {Klapwijk}}]{vercruyssen2012}%
  \BibitemOpen
  \bibfield  {author} {\bibinfo {author} {\bibfnamefont {N.}~\bibnamefont
  {Vercruyssen}}, \bibinfo {author} {\bibfnamefont {T.~G.~A.}\ \bibnamefont
  {Verhagen}}, \bibinfo {author} {\bibfnamefont {M.~G.}\ \bibnamefont
  {Flokstra}}, \bibinfo {author} {\bibfnamefont {J.~P.}\ \bibnamefont
  {Pekola}},\ and\ \bibinfo {author} {\bibfnamefont {T.~M.}\ \bibnamefont
  {Klapwijk}},\ }\bibfield  {title} {\bibinfo {title} {Evanescent states and
  nonequilibrium in driven superconducting nanowires},\ }\href
  {https://doi.org/10.1103/PhysRevB.85.224503} {\bibfield  {journal} {\bibinfo
  {journal} {Physical Review B}\ }\textbf {\bibinfo {volume} {85}},\ \bibinfo
  {pages} {224503} (\bibinfo {year} {2012})}\BibitemShut {NoStop}%
\bibitem [{\citenamefont {Fu}(2010)}]{fu2010}%
  \BibitemOpen
  \bibfield  {author} {\bibinfo {author} {\bibfnamefont {L.}~\bibnamefont
  {Fu}},\ }\bibfield  {title} {\bibinfo {title} {Electron {{Teleportation}} via
  {{Majorana Bound States}} in a {{Mesoscopic Superconductor}}},\ }\href
  {https://doi.org/10.1103/PhysRevLett.104.056402} {\bibfield  {journal}
  {\bibinfo  {journal} {Physical Review Letters}\ }\textbf {\bibinfo {volume}
  {104}},\ \bibinfo {pages} {056402} (\bibinfo {year} {2010})}\BibitemShut
  {NoStop}%
\bibitem [{\citenamefont {Ulrich}\ and\ \citenamefont
  {Hassler}(2015)}]{ulrich2015}%
  \BibitemOpen
  \bibfield  {author} {\bibinfo {author} {\bibfnamefont {J.}~\bibnamefont
  {Ulrich}}\ and\ \bibinfo {author} {\bibfnamefont {F.}~\bibnamefont
  {Hassler}},\ }\bibfield  {title} {\bibinfo {title} {Majorana-assisted
  nonlocal electron transport through a floating topological superconductor},\
  }\href {https://doi.org/10.1103/PhysRevB.92.075443} {\bibfield  {journal}
  {\bibinfo  {journal} {Physical Review B}\ }\textbf {\bibinfo {volume} {92}},\
  \bibinfo {pages} {075443} (\bibinfo {year} {2015})}\BibitemShut {NoStop}%
\bibitem [{\citenamefont {Albrecht}\ \emph {et~al.}(2016)\citenamefont
  {Albrecht}, \citenamefont {Higginbotham}, \citenamefont {Madsen},
  \citenamefont {Kuemmeth}, \citenamefont {Jespersen}, \citenamefont
  {Nyg{\aa}rd}, \citenamefont {Krogstrup},\ and\ \citenamefont
  {Marcus}}]{albrecht2016}%
  \BibitemOpen
  \bibfield  {author} {\bibinfo {author} {\bibfnamefont {S.~M.}\ \bibnamefont
  {Albrecht}}, \bibinfo {author} {\bibfnamefont {A.~P.}\ \bibnamefont
  {Higginbotham}}, \bibinfo {author} {\bibfnamefont {M.}~\bibnamefont
  {Madsen}}, \bibinfo {author} {\bibfnamefont {F.}~\bibnamefont {Kuemmeth}},
  \bibinfo {author} {\bibfnamefont {T.~S.}\ \bibnamefont {Jespersen}}, \bibinfo
  {author} {\bibfnamefont {J.}~\bibnamefont {Nyg{\aa}rd}}, \bibinfo {author}
  {\bibfnamefont {P.}~\bibnamefont {Krogstrup}},\ and\ \bibinfo {author}
  {\bibfnamefont {C.~M.}\ \bibnamefont {Marcus}},\ }\bibfield  {title}
  {\bibinfo {title} {Exponential protection of zero modes in {{Majorana}}
  islands},\ }\href {https://doi.org/10.1038/nature17162} {\bibfield  {journal}
  {\bibinfo  {journal} {Nature}\ }\textbf {\bibinfo {volume} {531}},\ \bibinfo
  {pages} {206} (\bibinfo {year} {2016})}\BibitemShut {NoStop}%
\bibitem [{\citenamefont {Lai}\ \emph {et~al.}(2021)\citenamefont {Lai},
  \citenamefont {Sarma},\ and\ \citenamefont {Sau}}]{lai2021}%
  \BibitemOpen
  \bibfield  {author} {\bibinfo {author} {\bibfnamefont {Y.-H.}\ \bibnamefont
  {Lai}}, \bibinfo {author} {\bibfnamefont {S.~D.}\ \bibnamefont {Sarma}},\
  and\ \bibinfo {author} {\bibfnamefont {J.~D.}\ \bibnamefont {Sau}},\
  }\bibfield  {title} {\bibinfo {title} {Theory of {{Coulomb}} blockaded
  transport in realistic {{Majorana}} nanowires},\ }\href
  {https://doi.org/10.1103/PhysRevB.104.085403} {\bibfield  {journal} {\bibinfo
   {journal} {Physical Review B}\ }\textbf {\bibinfo {volume} {104}},\ \bibinfo
  {pages} {085403} (\bibinfo {year} {2021})}\BibitemShut {NoStop}%
\bibitem [{\citenamefont {Hao}\ \emph {et~al.}(2022)\citenamefont {Hao},
  \citenamefont {Zhang}, \citenamefont {Liu},\ and\ \citenamefont
  {Liu}}]{hao2022}%
  \BibitemOpen
  \bibfield  {author} {\bibinfo {author} {\bibfnamefont {Y.}~\bibnamefont
  {Hao}}, \bibinfo {author} {\bibfnamefont {G.}~\bibnamefont {Zhang}}, \bibinfo
  {author} {\bibfnamefont {D.}~\bibnamefont {Liu}},\ and\ \bibinfo {author}
  {\bibfnamefont {D.~E.}\ \bibnamefont {Liu}},\ }\bibfield  {title} {\bibinfo
  {title} {Double {{Fu-teleportation}} and anomalous {{Coulomb}} blockade in a
  {{Majorana-hosted}} superconducting island},\ }\href
  {https://doi.org/10.1038/s41467-022-34437-9} {\bibfield  {journal} {\bibinfo
  {journal} {Nature Communications}\ }\textbf {\bibinfo {volume} {13}},\
  \bibinfo {pages} {6699} (\bibinfo {year} {2022})}\BibitemShut {NoStop}%
\bibitem [{\citenamefont {Souto}\ \emph {et~al.}(2022)\citenamefont {Souto},
  \citenamefont {Wauters}, \citenamefont {Flensberg}, \citenamefont {Leijnse},\
  and\ \citenamefont {Burrello}}]{souto2022}%
  \BibitemOpen
  \bibfield  {author} {\bibinfo {author} {\bibfnamefont {R.~S.}\ \bibnamefont
  {Souto}}, \bibinfo {author} {\bibfnamefont {M.~M.}\ \bibnamefont {Wauters}},
  \bibinfo {author} {\bibfnamefont {K.}~\bibnamefont {Flensberg}}, \bibinfo
  {author} {\bibfnamefont {M.}~\bibnamefont {Leijnse}},\ and\ \bibinfo {author}
  {\bibfnamefont {M.}~\bibnamefont {Burrello}},\ }\bibfield  {title} {\bibinfo
  {title} {Multiterminal transport spectroscopy of subgap states in
  {{Coulomb-blockaded}} superconductors},\ }\href
  {https://doi.org/10.1103/PhysRevB.106.235425} {\bibfield  {journal} {\bibinfo
   {journal} {Physical Review B}\ }\textbf {\bibinfo {volume} {106}},\ \bibinfo
  {pages} {235425} (\bibinfo {year} {2022})}\BibitemShut {NoStop}%
\bibitem [{\citenamefont {Valentini}\ \emph {et~al.}(2022)\citenamefont
  {Valentini}, \citenamefont {Borovkov}, \citenamefont {Prada}, \citenamefont
  {{Mart{\'i}-S{\'a}nchez}}, \citenamefont {Botifoll}, \citenamefont {Hofmann},
  \citenamefont {Arbiol}, \citenamefont {Aguado}, \citenamefont {{San-Jose}},\
  and\ \citenamefont {Katsaros}}]{valentini2022}%
  \BibitemOpen
  \bibfield  {author} {\bibinfo {author} {\bibfnamefont {M.}~\bibnamefont
  {Valentini}}, \bibinfo {author} {\bibfnamefont {M.}~\bibnamefont {Borovkov}},
  \bibinfo {author} {\bibfnamefont {E.}~\bibnamefont {Prada}}, \bibinfo
  {author} {\bibfnamefont {S.}~\bibnamefont {{Mart{\'i}-S{\'a}nchez}}},
  \bibinfo {author} {\bibfnamefont {M.}~\bibnamefont {Botifoll}}, \bibinfo
  {author} {\bibfnamefont {A.}~\bibnamefont {Hofmann}}, \bibinfo {author}
  {\bibfnamefont {J.}~\bibnamefont {Arbiol}}, \bibinfo {author} {\bibfnamefont
  {R.}~\bibnamefont {Aguado}}, \bibinfo {author} {\bibfnamefont
  {P.}~\bibnamefont {{San-Jose}}},\ and\ \bibinfo {author} {\bibfnamefont
  {G.}~\bibnamefont {Katsaros}},\ }\bibfield  {title} {\bibinfo {title}
  {Majorana-like {{Coulomb}} spectroscopy in the absence of zero-bias peaks},\
  }\href {https://doi.org/10.1038/s41586-022-05382-w} {\bibfield  {journal}
  {\bibinfo  {journal} {Nature}\ }\textbf {\bibinfo {volume} {612}},\ \bibinfo
  {pages} {442} (\bibinfo {year} {2022})}\BibitemShut {NoStop}%
\bibitem [{\citenamefont {Roddaro}\ \emph {et~al.}(2011)\citenamefont
  {Roddaro}, \citenamefont {Pescaglini}, \citenamefont {Ercolani},
  \citenamefont {Sorba}, \citenamefont {Giazotto},\ and\ \citenamefont
  {Beltram}}]{roddaro2011}%
  \BibitemOpen
  \bibfield  {author} {\bibinfo {author} {\bibfnamefont {S.}~\bibnamefont
  {Roddaro}}, \bibinfo {author} {\bibfnamefont {A.}~\bibnamefont {Pescaglini}},
  \bibinfo {author} {\bibfnamefont {D.}~\bibnamefont {Ercolani}}, \bibinfo
  {author} {\bibfnamefont {L.}~\bibnamefont {Sorba}}, \bibinfo {author}
  {\bibfnamefont {F.}~\bibnamefont {Giazotto}},\ and\ \bibinfo {author}
  {\bibfnamefont {F.}~\bibnamefont {Beltram}},\ }\bibfield  {title} {\bibinfo
  {title} {Hot-electron effects in {{InAs}} nanowire {{Josephson}} junctions},\
  }\href {https://doi.org/10.1007/s12274-010-0077-6} {\bibfield  {journal}
  {\bibinfo  {journal} {Nano Research}\ }\textbf {\bibinfo {volume} {4}},\
  \bibinfo {pages} {259} (\bibinfo {year} {2011})}\BibitemShut {NoStop}%
\bibitem [{\citenamefont {Bubis}(2021)}]{bubis2021}%
  \BibitemOpen
  \bibfield  {author} {\bibinfo {author} {\bibfnamefont {A.~V.}\ \bibnamefont
  {Bubis}},\ }\bibfield  {title} {\bibinfo {title} {Thermal conductance and
  nonequilibrium superconductivity in a diffusive {{NSN}} wire probed by shot
  noise},\ }\href {https://doi.org/10.1103/PhysRevB.104.125409} {\bibfield
  {journal} {\bibinfo  {journal} {Physical Review B}\ }\textbf {\bibinfo
  {volume} {104}},\ \bibinfo {pages} {125409} (\bibinfo {year}
  {2021})}\BibitemShut {NoStop}%
\bibitem [{\citenamefont {Liu}\ \emph {et~al.}(2023)\citenamefont {Liu},
  \citenamefont {Pan}, \citenamefont {Le}, \citenamefont {He}, \citenamefont
  {Jia}, \citenamefont {Zhu}, \citenamefont {Yang}, \citenamefont {Lyu},
  \citenamefont {Liu}, \citenamefont {Shen}, \citenamefont {Zhao},
  \citenamefont {Lu},\ and\ \citenamefont {Qu}}]{liu2023}%
  \BibitemOpen
  \bibfield  {author} {\bibinfo {author} {\bibfnamefont {M.-L.}\ \bibnamefont
  {Liu}}, \bibinfo {author} {\bibfnamefont {D.}~\bibnamefont {Pan}}, \bibinfo
  {author} {\bibfnamefont {T.}~\bibnamefont {Le}}, \bibinfo {author}
  {\bibfnamefont {J.-B.}\ \bibnamefont {He}}, \bibinfo {author} {\bibfnamefont
  {Z.-M.}\ \bibnamefont {Jia}}, \bibinfo {author} {\bibfnamefont
  {S.}~\bibnamefont {Zhu}}, \bibinfo {author} {\bibfnamefont {G.}~\bibnamefont
  {Yang}}, \bibinfo {author} {\bibfnamefont {Z.-Z.}\ \bibnamefont {Lyu}},
  \bibinfo {author} {\bibfnamefont {G.-T.}\ \bibnamefont {Liu}}, \bibinfo
  {author} {\bibfnamefont {J.}~\bibnamefont {Shen}}, \bibinfo {author}
  {\bibfnamefont {J.-H.}\ \bibnamefont {Zhao}}, \bibinfo {author}
  {\bibfnamefont {L.}~\bibnamefont {Lu}},\ and\ \bibinfo {author}
  {\bibfnamefont {F.-M.}\ \bibnamefont {Qu}},\ }\bibfield  {title} {\bibinfo
  {title} {Gate-{{Tunable Negative Differential Conductance}} in {{Hybrid
  Semiconductor}}{\textendash}{{Superconductor Devices}}},\ }\href
  {https://doi.org/10.1088/0256-307X/40/6/067301} {\bibfield  {journal}
  {\bibinfo  {journal} {Chinese Physics Letters}\ }\textbf {\bibinfo {volume}
  {40}},\ \bibinfo {pages} {067301} (\bibinfo {year} {2023})}\BibitemShut
  {NoStop}%
\bibitem [{\citenamefont {Ibabe}\ \emph
  {et~al.}(2023{\natexlab{a}})\citenamefont {Ibabe}, \citenamefont {G{\'o}mez},
  \citenamefont {Steffensen}, \citenamefont {Kanne}, \citenamefont
  {Nyg{\aa}rd}, \citenamefont {Yeyati},\ and\ \citenamefont {Lee}}]{ibabe2023}%
  \BibitemOpen
  \bibfield  {author} {\bibinfo {author} {\bibfnamefont {A.}~\bibnamefont
  {Ibabe}}, \bibinfo {author} {\bibfnamefont {M.}~\bibnamefont {G{\'o}mez}},
  \bibinfo {author} {\bibfnamefont {G.~O.}\ \bibnamefont {Steffensen}},
  \bibinfo {author} {\bibfnamefont {T.}~\bibnamefont {Kanne}}, \bibinfo
  {author} {\bibfnamefont {J.}~\bibnamefont {Nyg{\aa}rd}}, \bibinfo {author}
  {\bibfnamefont {A.~L.}\ \bibnamefont {Yeyati}},\ and\ \bibinfo {author}
  {\bibfnamefont {E.~J.~H.}\ \bibnamefont {Lee}},\ }\bibfield  {title}
  {\bibinfo {title} {Joule spectroscopy of hybrid
  superconductor{\textendash}semiconductor nanodevices},\ }\href
  {https://doi.org/10.1038/s41467-023-38533-2} {\bibfield  {journal} {\bibinfo
  {journal} {Nature Communications}\ }\textbf {\bibinfo {volume} {14}},\
  \bibinfo {pages} {2873} (\bibinfo {year} {2023}{\natexlab{a}})}\BibitemShut
  {NoStop}%
\bibitem [{sup()}]{supplemental}%
  \BibitemOpen
  \href@noop {} {\bibinfo {title} {See {{Supplemental Materials}} at
  [{{URL}}], which includes {{Refs}}. [74-77]}}\BibitemShut {NoStop}%
\bibitem [{\citenamefont {Anthore}\ \emph {et~al.}(2003)\citenamefont
  {Anthore}, \citenamefont {Pothier},\ and\ \citenamefont
  {Esteve}}]{anthore2003}%
  \BibitemOpen
  \bibfield  {author} {\bibinfo {author} {\bibfnamefont {A.}~\bibnamefont
  {Anthore}}, \bibinfo {author} {\bibfnamefont {H.}~\bibnamefont {Pothier}},\
  and\ \bibinfo {author} {\bibfnamefont {D.}~\bibnamefont {Esteve}},\
  }\bibfield  {title} {\bibinfo {title} {Density of {{States}} in a
  {{Superconductor Carrying}} a {{Supercurrent}}},\ }\href
  {https://doi.org/10.1103/PhysRevLett.90.127001} {\bibfield  {journal}
  {\bibinfo  {journal} {Physical Review Letters}\ }\textbf {\bibinfo {volume}
  {90}},\ \bibinfo {pages} {127001} (\bibinfo {year} {2003})}\BibitemShut
  {NoStop}%
\bibitem [{\citenamefont {Sourribes}\ \emph {et~al.}(2013)\citenamefont
  {Sourribes}, \citenamefont {Isakov}, \citenamefont {Panfilova},\ and\
  \citenamefont {Warburton}}]{sourribes2013}%
  \BibitemOpen
  \bibfield  {author} {\bibinfo {author} {\bibfnamefont {M.~J.~L.}\
  \bibnamefont {Sourribes}}, \bibinfo {author} {\bibfnamefont {I.}~\bibnamefont
  {Isakov}}, \bibinfo {author} {\bibfnamefont {M.}~\bibnamefont {Panfilova}},\
  and\ \bibinfo {author} {\bibfnamefont {P.~A.}\ \bibnamefont {Warburton}},\
  }\bibfield  {title} {\bibinfo {title} {Minimization of the contact resistance
  between {{InAs}} nanowires and metallic contacts},\ }\href
  {https://doi.org/10.1088/0957-4484/24/4/045703} {\bibfield  {journal}
  {\bibinfo  {journal} {Nanotechnology}\ }\textbf {\bibinfo {volume} {24}},\
  \bibinfo {pages} {045703} (\bibinfo {year} {2013})}\BibitemShut {NoStop}%
\bibitem [{\citenamefont {Nazarov}(1999)}]{nazarov1999}%
  \BibitemOpen
  \bibfield  {author} {\bibinfo {author} {\bibfnamefont {Y.~V.}\ \bibnamefont
  {Nazarov}},\ }\bibfield  {title} {\bibinfo {title} {Coulomb {{Blockade}}
  without {{Tunnel Junctions}}},\ }\href
  {https://doi.org/10.1103/PhysRevLett.82.1245} {\bibfield  {journal} {\bibinfo
   {journal} {Physical Review Letters}\ }\textbf {\bibinfo {volume} {82}},\
  \bibinfo {pages} {1245} (\bibinfo {year} {1999})}\BibitemShut {NoStop}%
\bibitem [{\citenamefont {Golubev}\ and\ \citenamefont
  {Zaikin}(2001)}]{golubev2001}%
  \BibitemOpen
  \bibfield  {author} {\bibinfo {author} {\bibfnamefont {D.~S.}\ \bibnamefont
  {Golubev}}\ and\ \bibinfo {author} {\bibfnamefont {A.~D.}\ \bibnamefont
  {Zaikin}},\ }\bibfield  {title} {\bibinfo {title} {Coulomb {{Interaction}}
  and {{Quantum Transport}} through a {{Coherent Scatterer}}},\ }\href
  {https://doi.org/10.1103/PhysRevLett.86.4887} {\bibfield  {journal} {\bibinfo
   {journal} {Physical Review Letters}\ }\textbf {\bibinfo {volume} {86}},\
  \bibinfo {pages} {4887} (\bibinfo {year} {2001})}\BibitemShut {NoStop}%
\bibitem [{\citenamefont {Yeyati}\ \emph {et~al.}(2001)\citenamefont {Yeyati},
  \citenamefont {{Martin-Rodero}}, \citenamefont {Esteve},\ and\ \citenamefont
  {Urbina}}]{yeyati2001}%
  \BibitemOpen
  \bibfield  {author} {\bibinfo {author} {\bibfnamefont {A.~L.}\ \bibnamefont
  {Yeyati}}, \bibinfo {author} {\bibfnamefont {A.}~\bibnamefont
  {{Martin-Rodero}}}, \bibinfo {author} {\bibfnamefont {D.}~\bibnamefont
  {Esteve}},\ and\ \bibinfo {author} {\bibfnamefont {C.}~\bibnamefont
  {Urbina}},\ }\bibfield  {title} {\bibinfo {title} {Direct {{Link}} between
  {{Coulomb Blockade}} and {{Shot Noise}} in a {{Quantum-Coherent
  Structure}}},\ }\href {https://doi.org/10.1103/PhysRevLett.87.046802}
  {\bibfield  {journal} {\bibinfo  {journal} {Physical Review Letters}\
  }\textbf {\bibinfo {volume} {87}},\ \bibinfo {pages} {046802} (\bibinfo
  {year} {2001})}\BibitemShut {NoStop}%
\bibitem [{\citenamefont {Nagaev}(1992)}]{nagaev1992}%
  \BibitemOpen
  \bibfield  {author} {\bibinfo {author} {\bibfnamefont {K.}~\bibnamefont
  {Nagaev}},\ }\bibfield  {title} {\bibinfo {title} {On the shot noise in dirty
  metal contacts},\ }\href {https://doi.org/10.1016/0375-9601(92)90814-3}
  {\bibfield  {journal} {\bibinfo  {journal} {Physics Letters A}\ }\textbf
  {\bibinfo {volume} {169}},\ \bibinfo {pages} {103} (\bibinfo {year}
  {1992})}\BibitemShut {NoStop}%
\bibitem [{\citenamefont {Beenakker}\ and\ \citenamefont
  {Buttiker}(1992)}]{beenakker1992}%
  \BibitemOpen
  \bibfield  {author} {\bibinfo {author} {\bibfnamefont {C.~W.~J.}\
  \bibnamefont {Beenakker}}\ and\ \bibinfo {author} {\bibfnamefont
  {M.}~\bibnamefont {Buttiker}},\ }\bibfield  {title} {\bibinfo {title}
  {Suppression of shot noise in metallic diffusive conductors},\ }\href
  {https://doi.org/10.1103/PhysRevB.46.1889} {\bibfield  {journal} {\bibinfo
  {journal} {Physical Review B}\ }\textbf {\bibinfo {volume} {46}},\ \bibinfo
  {pages} {1889} (\bibinfo {year} {1992})}\BibitemShut {NoStop}%
\bibitem [{\citenamefont {{E S Tikhonov}}\ \emph {et~al.}(2016)\citenamefont
  {{E S Tikhonov}}, \citenamefont {{D V Shovkun}}, \citenamefont {{D
  Ercolani}}, \citenamefont {{F Rossella}}, \citenamefont {{M Rocci}},
  \citenamefont {{L Sorba}}, \citenamefont {{S Roddaro}},\ and\ \citenamefont
  {{V S Khrapai}}}]{tikhonov2016a}%
  \BibitemOpen
  \bibfield  {author} {\bibinfo {author} {\bibnamefont {{E S Tikhonov}}},
  \bibinfo {author} {\bibnamefont {{D V Shovkun}}}, \bibinfo {author}
  {\bibnamefont {{D Ercolani}}}, \bibinfo {author} {\bibnamefont {{F
  Rossella}}}, \bibinfo {author} {\bibnamefont {{M Rocci}}}, \bibinfo {author}
  {\bibnamefont {{L Sorba}}}, \bibinfo {author} {\bibnamefont {{S Roddaro}}},\
  and\ \bibinfo {author} {\bibnamefont {{V S Khrapai}}},\ }\bibfield  {title}
  {\bibinfo {title} {Local noise in a diffusive conductor},\ }\href
  {https://doi.org/10.1038/srep30621} {\bibfield  {journal} {\bibinfo
  {journal} {Scientific Reports}\ }\textbf {\bibinfo {volume} {6}},\ \bibinfo
  {pages} {30621} (\bibinfo {year} {2016})}\BibitemShut {NoStop}%
\bibitem [{\citenamefont {Tikhonov}(2016)}]{tikhonov2016b}%
  \BibitemOpen
  \bibfield  {author} {\bibinfo {author} {\bibfnamefont {E.~S.}\ \bibnamefont
  {Tikhonov}},\ }\bibfield  {title} {\bibinfo {title} {Noise thermometry
  applied to thermoelectric measurements in {{InAs}} nanowires},\ }\href
  {https://doi.org/10.1088/0268-1242/31/10/104001} {\bibfield  {journal}
  {\bibinfo  {journal} {Semicond. Sci. Technol.}\ }\textbf {\bibinfo {volume}
  {31}},\ \bibinfo {pages} {104001} (\bibinfo {year} {2016})}\BibitemShut
  {NoStop}%
\bibitem [{\citenamefont {De~Jong}\ and\ \citenamefont
  {Beenakker}(1996)}]{dejong1996}%
  \BibitemOpen
  \bibfield  {author} {\bibinfo {author} {\bibfnamefont {M.}~\bibnamefont
  {De~Jong}}\ and\ \bibinfo {author} {\bibfnamefont {C.}~\bibnamefont
  {Beenakker}},\ }\bibfield  {title} {\bibinfo {title} {Semiclassical theory of
  shot noise in mesoscopic conductors},\ }\href
  {https://doi.org/10.1016/0378-4371(96)00068-4} {\bibfield  {journal}
  {\bibinfo  {journal} {Physica A: Statistical Mechanics and its Applications}\
  }\textbf {\bibinfo {volume} {230}},\ \bibinfo {pages} {219} (\bibinfo {year}
  {1996})}\BibitemShut {NoStop}%
\bibitem [{\citenamefont {Pinsolle}\ \emph {et~al.}(2016)\citenamefont
  {Pinsolle}, \citenamefont {Rousseau}, \citenamefont {Lupien},\ and\
  \citenamefont {Reulet}}]{pinsolle2016}%
  \BibitemOpen
  \bibfield  {author} {\bibinfo {author} {\bibfnamefont {E.}~\bibnamefont
  {Pinsolle}}, \bibinfo {author} {\bibfnamefont {A.}~\bibnamefont {Rousseau}},
  \bibinfo {author} {\bibfnamefont {C.}~\bibnamefont {Lupien}},\ and\ \bibinfo
  {author} {\bibfnamefont {B.}~\bibnamefont {Reulet}},\ }\bibfield  {title}
  {\bibinfo {title} {Direct measurement of the electron energy relaxation
  dynamics in metallic wires},\ }\href
  {https://doi.org/10.1103/PhysRevLett.116.236601} {\bibfield  {journal}
  {\bibinfo  {journal} {Physical Review Letters}\ }\textbf {\bibinfo {volume}
  {116}},\ \bibinfo {pages} {236601} (\bibinfo {year} {2016})}\BibitemShut
  {NoStop}%
\bibitem [{\citenamefont {V{\"a}yrynen}\ \emph {et~al.}(2021)\citenamefont
  {V{\"a}yrynen}, \citenamefont {Pikulin},\ and\ \citenamefont
  {Lutchyn}}]{vayrynen2021}%
  \BibitemOpen
  \bibfield  {author} {\bibinfo {author} {\bibfnamefont {J.~I.}\ \bibnamefont
  {V{\"a}yrynen}}, \bibinfo {author} {\bibfnamefont {D.~I.}\ \bibnamefont
  {Pikulin}},\ and\ \bibinfo {author} {\bibfnamefont {R.~M.}\ \bibnamefont
  {Lutchyn}},\ }\bibfield  {title} {\bibinfo {title} {Majorana signatures in
  charge transport through a topological superconducting double-island
  system},\ }\href {https://doi.org/10.1103/PhysRevB.103.205427} {\bibfield
  {journal} {\bibinfo  {journal} {Physical Review B}\ }\textbf {\bibinfo
  {volume} {103}},\ \bibinfo {pages} {205427} (\bibinfo {year}
  {2021})}\BibitemShut {NoStop}%
\bibitem [{\citenamefont {Ibabe}\ \emph
  {et~al.}(2023{\natexlab{b}})\citenamefont {Ibabe}, \citenamefont
  {Steffensen}, \citenamefont {Casal}, \citenamefont {Gomez}, \citenamefont
  {Kanne}, \citenamefont {Nygard}, \citenamefont {Yeyati},\ and\ \citenamefont
  {Lee}}]{ibabe2023a}%
  \BibitemOpen
  \bibfield  {author} {\bibinfo {author} {\bibfnamefont {A.}~\bibnamefont
  {Ibabe}}, \bibinfo {author} {\bibfnamefont {G.~O.}\ \bibnamefont
  {Steffensen}}, \bibinfo {author} {\bibfnamefont {I.}~\bibnamefont {Casal}},
  \bibinfo {author} {\bibfnamefont {M.}~\bibnamefont {Gomez}}, \bibinfo
  {author} {\bibfnamefont {T.}~\bibnamefont {Kanne}}, \bibinfo {author}
  {\bibfnamefont {J.}~\bibnamefont {Nygard}}, \bibinfo {author} {\bibfnamefont
  {A.~L.}\ \bibnamefont {Yeyati}},\ and\ \bibinfo {author} {\bibfnamefont
  {E.~J.~H.}\ \bibnamefont {Lee}},\ }\bibfield  {title} {\bibinfo {title} {Heat
  dissipation mechanisms in hybrid superconductor-semiconductor devices
  revealed by {{Joule}} spectroscopy},\ }\href
  {https://doi.org/10.48550/arXiv.2311.13229} {\bibfield  {journal} {\bibinfo
  {journal} {arXiv:2311.13229 [cond-mat.mes-hall]}\ } (\bibinfo {year}
  {2023}{\natexlab{b}})}\BibitemShut {NoStop}%
\bibitem [{\citenamefont {Ruhstorfer}\ \emph {et~al.}(2021)\citenamefont
  {Ruhstorfer}, \citenamefont {Lang}, \citenamefont {Matich}, \citenamefont
  {D{\"o}blinger}, \citenamefont {Riedl}, \citenamefont {Finley},\ and\
  \citenamefont {Koblm{\"u}ller}}]{ruhstorfer2021}%
  \BibitemOpen
  \bibfield  {author} {\bibinfo {author} {\bibfnamefont {D.}~\bibnamefont
  {Ruhstorfer}}, \bibinfo {author} {\bibfnamefont {A.}~\bibnamefont {Lang}},
  \bibinfo {author} {\bibfnamefont {S.}~\bibnamefont {Matich}}, \bibinfo
  {author} {\bibfnamefont {M.}~\bibnamefont {D{\"o}blinger}}, \bibinfo {author}
  {\bibfnamefont {H.}~\bibnamefont {Riedl}}, \bibinfo {author} {\bibfnamefont
  {J.~J.}\ \bibnamefont {Finley}},\ and\ \bibinfo {author} {\bibfnamefont
  {G.}~\bibnamefont {Koblm{\"u}ller}},\ }\bibfield  {title} {\bibinfo {title}
  {Growth dynamics and compositional structure in periodic {{InAsSb}} nanowire
  arrays on {{Si}} (111) grown by selective area molecular beam epitaxy},\
  }\href {https://doi.org/10.1088/1361-6528/abcdca} {\bibfield  {journal}
  {\bibinfo  {journal} {Nanotechnology}\ }\textbf {\bibinfo {volume} {32}},\
  \bibinfo {pages} {135604} (\bibinfo {year} {2021})}\BibitemShut {NoStop}%
\bibitem [{\citenamefont {Del~Giudice}\ \emph {et~al.}(2020)\citenamefont
  {Del~Giudice}, \citenamefont {Becker}, \citenamefont {De~Rose}, \citenamefont
  {D{\"o}blinger}, \citenamefont {Ruhstorfer}, \citenamefont {Suomenniemi},
  \citenamefont {Treu}, \citenamefont {Riedl}, \citenamefont {Finley},\ and\
  \citenamefont {Koblm{\"u}ller}}]{delgiudice2020}%
  \BibitemOpen
  \bibfield  {author} {\bibinfo {author} {\bibfnamefont {F.}~\bibnamefont
  {Del~Giudice}}, \bibinfo {author} {\bibfnamefont {J.}~\bibnamefont {Becker}},
  \bibinfo {author} {\bibfnamefont {C.}~\bibnamefont {De~Rose}}, \bibinfo
  {author} {\bibfnamefont {M.}~\bibnamefont {D{\"o}blinger}}, \bibinfo {author}
  {\bibfnamefont {D.}~\bibnamefont {Ruhstorfer}}, \bibinfo {author}
  {\bibfnamefont {L.}~\bibnamefont {Suomenniemi}}, \bibinfo {author}
  {\bibfnamefont {J.}~\bibnamefont {Treu}}, \bibinfo {author} {\bibfnamefont
  {H.}~\bibnamefont {Riedl}}, \bibinfo {author} {\bibfnamefont {J.~J.}\
  \bibnamefont {Finley}},\ and\ \bibinfo {author} {\bibfnamefont
  {G.}~\bibnamefont {Koblm{\"u}ller}},\ }\bibfield  {title} {\bibinfo {title}
  {Ultrathin catalyst-free {{InAs}} nanowires on silicon with distinct {{1D}}
  sub-band transport properties},\ }\href {https://doi.org/10.1039/D0NR05666A}
  {\bibfield  {journal} {\bibinfo  {journal} {Nanoscale}\ }\textbf {\bibinfo
  {volume} {12}},\ \bibinfo {pages} {21857} (\bibinfo {year}
  {2020})}\BibitemShut {NoStop}%
\bibitem [{\citenamefont {Rudolph}\ \emph {et~al.}(2013)\citenamefont
  {Rudolph}, \citenamefont {Funk}, \citenamefont {D{\"o}blinger}, \citenamefont
  {Mork{\"o}tter}, \citenamefont {Hertenberger}, \citenamefont {Schweickert},
  \citenamefont {Becker}, \citenamefont {Matich}, \citenamefont {Bichler},
  \citenamefont {Spirkoska}, \citenamefont {Zardo}, \citenamefont {Finley},
  \citenamefont {Abstreiter},\ and\ \citenamefont
  {Koblm{\"u}ller}}]{rudolph2013}%
  \BibitemOpen
  \bibfield  {author} {\bibinfo {author} {\bibfnamefont {D.}~\bibnamefont
  {Rudolph}}, \bibinfo {author} {\bibfnamefont {S.}~\bibnamefont {Funk}},
  \bibinfo {author} {\bibfnamefont {M.}~\bibnamefont {D{\"o}blinger}}, \bibinfo
  {author} {\bibfnamefont {S.}~\bibnamefont {Mork{\"o}tter}}, \bibinfo {author}
  {\bibfnamefont {S.}~\bibnamefont {Hertenberger}}, \bibinfo {author}
  {\bibfnamefont {L.}~\bibnamefont {Schweickert}}, \bibinfo {author}
  {\bibfnamefont {J.}~\bibnamefont {Becker}}, \bibinfo {author} {\bibfnamefont
  {S.}~\bibnamefont {Matich}}, \bibinfo {author} {\bibfnamefont
  {M.}~\bibnamefont {Bichler}}, \bibinfo {author} {\bibfnamefont
  {D.}~\bibnamefont {Spirkoska}}, \bibinfo {author} {\bibfnamefont
  {I.}~\bibnamefont {Zardo}}, \bibinfo {author} {\bibfnamefont {J.~J.}\
  \bibnamefont {Finley}}, \bibinfo {author} {\bibfnamefont {G.}~\bibnamefont
  {Abstreiter}},\ and\ \bibinfo {author} {\bibfnamefont {G.}~\bibnamefont
  {Koblm{\"u}ller}},\ }\bibfield  {title} {\bibinfo {title} {Spontaneous
  {{Alloy Composition Ordering}} in {{GaAs-AlGaAs Core}}{\textendash}{{Shell
  Nanowires}}},\ }\href {https://doi.org/10.1021/nl3046816} {\bibfield
  {journal} {\bibinfo  {journal} {Nano Letters}\ }\textbf {\bibinfo {volume}
  {13}},\ \bibinfo {pages} {1522} (\bibinfo {year} {2013})}\BibitemShut
  {NoStop}%
\bibitem [{\citenamefont {Tikhonov}\ \emph {et~al.}(2014)\citenamefont
  {Tikhonov}, \citenamefont {Melnikov}, \citenamefont {Shovkun}, \citenamefont
  {Sorba}, \citenamefont {Biasiol},\ and\ \citenamefont
  {Khrapai}}]{tikhonov2014a}%
  \BibitemOpen
  \bibfield  {author} {\bibinfo {author} {\bibfnamefont {E.~S.}\ \bibnamefont
  {Tikhonov}}, \bibinfo {author} {\bibfnamefont {M.~{\relax Yu}.}\ \bibnamefont
  {Melnikov}}, \bibinfo {author} {\bibfnamefont {D.~V.}\ \bibnamefont
  {Shovkun}}, \bibinfo {author} {\bibfnamefont {L.}~\bibnamefont {Sorba}},
  \bibinfo {author} {\bibfnamefont {G.}~\bibnamefont {Biasiol}},\ and\ \bibinfo
  {author} {\bibfnamefont {V.~S.}\ \bibnamefont {Khrapai}},\ }\bibfield
  {title} {\bibinfo {title} {Nonlinear transport and noise thermometry in
  quasiclassical ballistic point contacts},\ }\href
  {https://doi.org/10.1103/PhysRevB.90.161405} {\bibfield  {journal} {\bibinfo
  {journal} {Physical Review B}\ }\textbf {\bibinfo {volume} {90}},\ \bibinfo
  {pages} {161405} (\bibinfo {year} {2014})}\BibitemShut {NoStop}%
\end{thebibliography}

%

\end{document}


\title{Fate of the superconducting state in floating islands of hybrid nanowire devices. Supplemental Materials.}
\author{E.V.~Shpagina}\affiliation{Osipyan Institute of Solid State Physics, Russian Academy of
Sciences, 142432 Chernogolovka, Russian Federation}
\affiliation{National Research University Higher School of Economics, 20 Myasnitskaya Street, 101000 Moscow, Russian Federation}

\author{E.S.~Tikhonov}\affiliation{Osipyan Institute of Solid State Physics, Russian Academy of
Sciences, 142432 Chernogolovka, Russian Federation}
\affiliation{National Research University Higher School of Economics, 20 Myasnitskaya Street, 101000 Moscow, Russian Federation}
\author{D.~Ruhstorfer}
\affiliation{Walter Schottky Institut, Physik Department, and Center for Nanotechnology and Nanomaterials, Technische Universit\"{a}t M\"{u}nchen, Am Coulombwall 4, Garching 85748, Germany}
\author{G.~Koblm\"{u}ller}
\affiliation{Walter Schottky Institut, Physik Department, and Center for Nanotechnology and Nanomaterials, Technische Universit\"{a}t M\"{u}nchen, Am Coulombwall 4, Garching 85748, Germany}
\author{V.S.~Khrapai}
\affiliation{Osipyan Institute of Solid State Physics, Russian Academy of
Sciences, 142432 Chernogolovka, Russian Federation}
\affiliation{National Research University Higher School of Economics, 20 Myasnitskaya Street, 101000 Moscow, Russian Federation}
\email{dick@issp.ac.ru}

\maketitle

\section{Nanowire growth}

The InAs-Al nanowire (NW) array is grown by molecular beam epitaxy (MBE) via a position-controlled selective area growth (SAG) approach on SiO$_2$/Si(111) substrates~\cite{ruhstorfer2021,delgiudice2020}. First, the InAs NW cores are grown for 50\,min in a non-catalytic (entirely droplet-free) growth mode at a temperature of 540~$^\circ$C under high V/III ratio. For the given SAG pattern (pitch of 2~$\mu$m, see Fig. 1a in the main text) the resulting NW-length and -diameter of the InAs NWs is about $\approx4\pm0.5\,\mu$m and $\approx160\pm10$\,nm, respectively. Subsequently, the as-grown InAs NW array is cooled down under arsenic (As$_4$) atmosphere to 300~$^\circ$C. Once this temperature is stabilized, the As$_4$-supply is switched off and the MBE chamber pumped until a background pressure of $9\times10^{-11}$\,mbar is reached. Afterwards the sample is incrementally heated up to $\sim450\,^\circ$C, in order to desorb any residual As$_4$ from the NW sidewall surfaces. In the next step, the NW array is once again cooled down, but to a very low temperature of -40~$^\circ$C, at which a $\approx$50-nm thick Al shell is conformally deposited in-situ onto the InAs NWs under constant substrate rotation (20 rpm). Hereby, a nominal Al-flux of 1.74~\AA/s is employed, as supplied from a conventional Al Knudsen effusion cell. Due to the NW-geometry and the orientation of the incident Al-flux, the growth rate on the NW-sidewalls is, however, by a factor of $\sim5$ less than the nominal, planar Al-rate~\cite{rudolph2013}. The Al shell layer is deposited in 12 consecutive steps (2 min-long Al deposition with 60-min pauses in between), to avoid any excessive heating of the NW array imposed by the hot Al Knudsen cell (cell temperature of 1200~$^\circ$C). This way the substrate temperature is kept constant between -40~$^\circ$C and -20~$^\circ$C upon completing the entire Al shell growth.

\section{Theoretical framework}

\subsection{Depairing factor}
Following previous work on full-shell nanowires~\cite{vaitiekenas2020,ibabe2023}, we employ a hollow cylinder model for the Al shell (epi-Al) in arbitrary magnetic field, taking into account the finite shell thickness ($t$).
We assume that the thickness of the superconducting film is less than the correlation length ($t<\xi_0$), so that the depairing factor  will include only the radius-averaged square of the superconducting velocity~\cite{anthore2003}:
$ \Gamma = \frac{\hbar}{2D} \langle \vec v_S ^2 \rangle $,
where $ \vec v_S  = D \left(
\vec{\nabla} \varphi - \frac{2e}{\hbar} \vec{A}
\right)$.
We admit that the magnetic field is not directed strictly parallel to the nanowire, but it has a small perpendicular component.
Therefore, in a cylindrical coordinate system ($z$,$\rho$,$\alpha$), we determine a calibration of the vector potential as $(A_z,A_\rho,A_\alpha)=(B_{\perp}\rho \sin \alpha,0,B_{\parallel} \rho/2)$. Without current, the phase gradient has only the azimuthal nonzero component and the integral of its gradient along a closed contour inside the cylinder is a multiple of $2\pi$:
\begin{equation*}
	\oint \vec{\nabla} \varphi d \vec{s} = 2\pi n, ~~~~~~ \vec{\nabla} \varphi = \frac{n}{\rho} \vec{\alpha}.
\end{equation*}
where $n$ is the winding number, $\vec{\alpha}$ is the unit vector, $\rho$ is intermediate radius of the shell.
The averaging operation is carried out over the entire cross-sectional area of the shell. The thickness of the shell is $t=\rho_\mathrm{o} - \rho_\mathrm{i}$, where $\rho_\mathrm{i}$ and $\rho_\mathrm{o}$ are the  inner and outer radii of the shell:
\begin{equation*}
	\langle \vec v_S^2 \rangle = \frac{\int_{\rho_\mathrm{i}}^{\rho_\mathrm{o}} \int_{0}^{2\pi}
	\langle \vec v_S^2 \rangle
	\rho d\rho d\alpha  }{\pi (\rho_\mathrm{o}^2 - \rho_\mathrm{i}^2 ) }
\end{equation*}
We find the depairing factor as a function of the winding number and $B$-field, which can be divided into parallel and perpendicular components:
 $\Gamma(B,n) = \Gamma_{\perp} + \Gamma_{\parallel}$:
\begin{align*}
	\Gamma_{\parallel} &= \frac{\hbar D}{2 \rho_+^2}
\left[
\left(n-\frac{\Phi}{\Phi_0}\right)^2
+ n^2 
\left(
\frac{\rho_+^2}{\rho_-^2}
\ln \frac{\rho_\mathrm{o}}{\rho_\mathrm{i}} - 1
\right)
\right], \\
\Gamma_{\perp} &= \frac{\hbar D}{\Phi_0^2} (\pi B_\perp \rho_+)^2
\end{align*}
where $D$ is the diffusion coefficient, $\rho_\pm^2 = (\rho_\mathrm{o}^2 \pm \rho_\mathrm{i}^2)/2 $, $\Phi=\pi \rho_+^2 B_\parallel$ and $n=0,1,2$ is the number of the Little-Parks lobe, $\Phi_0 = h/2e$.
The depairing factor for the device IIA in the cooldown 2 is shown in the Fig.~1e of the manuscript.

\subsection{Critical temperature}
The dotted line fit on the Fig.~1c of the manuscript is the solution of well-known Abrikosov-Gorkov equation for the critical temperature of a superconductor with finite $\Gamma$:
\begin{align}
	\ln \left( \frac{T_\mathrm{c}}{T_\mathrm{c}^0} \right) = \Psi \left(\frac12 \right) - \Psi \left(\frac12 + \frac{\Gamma(B,n) }{2\pi T_\mathrm{c}} \right) \label{AG_eq}
\end{align}
where $\Psi$ is the digamma function, $T_\mathrm{c}^0\approx 1.23$\,K is the $B=0$ critical temperature, directly measured in  the device IA.

\subsection{Critical voltage}

\begin{figure}[t!]
	\begin{center}
		\vspace{0mm}
		\includegraphics[width=0.8\linewidth]{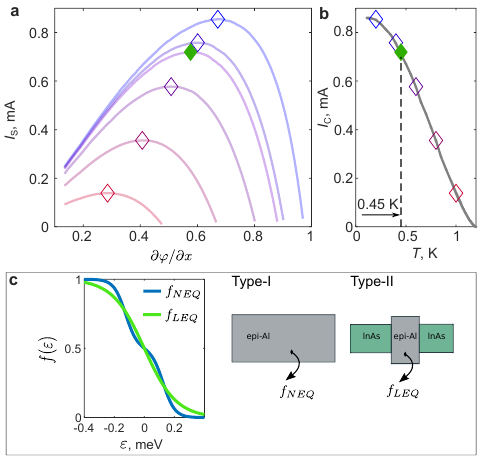}
	\end{center}
	\caption{(a)  The supercurrent as a function of phase gradient at several temperatures is plotted for the IA device. (b) The critical current for the IA device. (c) Distribution functions in the middle of the epi-Al for both types of devices and nanowire layouts. } 
	\label{FigS1}
\end{figure}

In our theoretical calculations we use the Usadel formalism (dirty limit) following Ref.~\cite{anthore2003}, supplementing it with non-equilibrium electronic energy distribution (EED) when necessary. A full treatment of the non-equilibrium problem has been carried out in Ref.~\cite{keizer2006}, where a self-consistent solution was found for the conversion of quasiparticle current into Cooper pair current near the NS interfaces. Here we solve a much simpler problem and only find the order parameter in the depth of the epi-Al, at distances much larger than $\xi$ from the NS interfaces. Under these conditions, the charge-mode quasiparticle non-equilibrium is absent and the order parameter is determined by the longitudinal (heat-mode or energy-mode) non-equilibrium, which is independent of the coordinate~\cite{keizer2006,snyman2009}. We also neglect the depairing factor due to finite supercurrent, which is justified by the fact that typical currents in the NSN transport regime are a factor of 4 smaller compared to the thermodynamical critical current (see below). Hence, the Usadel equation reads:
 \begin{align}
 	E + i \Gamma \cos \theta = i \Delta \frac{\cos \theta}{\sin \theta} \label{uzadel}
 \end{align}
where $\theta(E)$ is a complex energy-dependent pairing angle. The order parameter is found from self-consistent BCS equation, which can be written in the form:
\begin{align}
\Delta =\ln^{-1} \left( \frac{2 E_\mathrm{D}}{\Delta_0} \right) 
\int_0^{E_D} dE f_\mathrm{L}(E) Im (\sin \theta) \label{BCS}
\end{align}
where $E_\mathrm{D}\equiv k_\mathrm{B}\Theta_\mathrm{D}$ is the Debye energy of aluminum ($\Theta_\mathrm{D}=429\,$K), $\Delta_0$ is the superconducting gap in the zero temperature and $B=0$ limit. The longitudinal component of the non-equilibrium EED is even function of energy defined as  $f_\mathrm{L}(E) =1-f(E)-f(-E)$ ~\cite{keizer2006}. In type-I devices the EED has a double-step shape $f_\mathrm{NEQ}(E)=\left[f_0(E-V/2,T_\mathrm{0})+f_0(E+V/2,T_\mathrm{0})\right]/2$, so that $f_\mathrm{L}(E) = \frac12 \left[ \tanh\left( \frac{E+eV}{2k_\mathrm{B} T_\mathrm{0}} \right) +  \tanh\left( \frac{E-eV}{2k_\mathrm{B}T_\mathrm{0}} \right) \right]$. Here $f_0(E,T_\mathrm{0})$ is the Fermi-Dirac EED at the base temperature $T_\mathrm{0}$. In type-II devices the local equilibrium EED $f_0(E,T_\mathrm{e})$ with the electronic temperature $T_\mathrm{e}$ is assumed, so that $f_\mathrm{L}(E)=\tanh\left( E/2k_\mathrm{B} T_\mathrm{e}\right)$. Corresponding EEDs in the center of the superconducting shell in type-I and type-II devices are shown in Fig.~\ref{FigS1}c for the case of $B\rightarrow0$ and $T_0=0.45$\,K, right before the collapse of the superconducting order parameter. 

The joint solution of equations (\ref{uzadel})-(\ref{BCS}) makes it possible to determine the order parameter for a given EED and $\Gamma$.  In type-I devices the solution $\Delta(V)$ is found. We pay attention to the bistability of solutions, which is detailed in previous works~\cite{keizer2006,snyman2009}. The procedure to find the critical voltage $V_\mathrm{C}$ is described in the main text. In type-II devices the procedure is simpler. Here $\Delta$ vanishes at $T_\mathrm{e} = T_\mathrm{c}(\Gamma)$ and the solution of Eqs. (\ref{uzadel})-(\ref{BCS}) is equivalent to that of equation (\ref{AG_eq}). 

\subsection{Critical current}

Within the framework of the Usadel formalism, it is possible to directly calculate the supercurrent of the epi-Al $I_\mathrm{S}= j \pi^2(\rho_\mathrm{o}-\rho_\mathrm{i})$. The supercurrent density is related to the charge density ($\rho_\mathrm{S}$) and the phase gradient as $\vec{j} = \rho_S D\vec{\nabla} \varphi $, where $ \rho_\mathrm{S} = e N(0) U_\mathrm{S} $ and $U_\mathrm{S}=\int_0^{\infty} dE f_\mathrm{L}(E) Im(\sin^2\theta)$. The pairing angle is obtained from the equation~(\ref{uzadel}) with $\Gamma = \frac{\hbar}{2D} \langle \vec v_S ^2 \rangle$.
The density of states is convenient to express via conductivity and diffusion coefficient $N(0)= \sigma/(e^2 D)$. We find the critical current in thermodynamic equilibrium and in the limit of $B=0$, so that the phase gradient is directed along the NW axis. Thus, the supercurrent reads:
\begin{align*}	
	I_\mathrm{S} = \frac{2\pi \sigma \rho_{-}^2}{e} \left(\frac{\partial \varphi}{\partial x}\right) 	\int_0^{\infty} dE f_\mathrm{L}(E) Im(\sin^2\theta) 
\end{align*}

Note that $2\pi \sigma \rho_{-}^2 = (R_\mathrm{sh}/L)^{-1}$, where $L$ is the length of the shell and $R_\mathrm{sh}$ is the shell resistance. The dependencies of $I_\mathrm{S}$ on $\partial \varphi/\partial x$ at various bath temperatures are displayed in Fig.~\ref{FigS1}a. For each temperature, the thermodynamical critical current $I_\mathrm{c}$ corresponds to the maximum of $I_\mathrm{S}$ and is marked by a diamond. The $T$-dependence of the $I_\mathrm{c}$ is plotted in Fig.~\ref{FigS1}b for the device IA. At the base temperature $T_0=0.45$\,K we find $I_\mathrm{c}\approx0.7$\,mA, which is higher than the experimental $I_\mathrm{c}$ measured in $B=0$. We attribute this discrepancy to the effect of finite interface transparency between the epi-Al and evap-Al layers. In the device IA $I_\mathrm{c}\approx0.35$\,mA and $R_\mathrm{int}\approx1.33\,\Omega$, whereas in the device IB $I_\mathrm{c}\approx0.49$\,mA and $R_\mathrm{int}\approx1.15\,\Omega$, supporting a correlation of $I_\mathrm{c}$ and $R_\mathrm{int}$.

\subsection{Consistency of numbers}

In our calculations we use a standard number for the density of states in aluminum,  $N(0)\approx2.2\times10^{47}\,\mathrm{J^{-1}m^{-3}}$. The diffusion coefficient $D=69\,\mathrm{cm^2/s}$ is obtained from the fit of $T_\mathrm{c}(B)$, which allows to compare a theoretical estimate of the shell resistance per length $R_\mathrm{sh}/L=1\,\Omega/\mu\mathrm{m}$ with the experimental data. In devices IA ($R_\mathrm{sh}\approx3.3\,\Omega$ and $L\approx3.1\,\mu$m) and IB ($R_\mathrm{sh}\approx2.86\,\Omega$ and $L\approx2.7\,\mu$m) we find similar values of $R_\mathrm{sh}/L\approx1.06\,\Omega/\mu\mathrm{m}$, which demonstrates a consistency of our approach.

\subsection{Electron-phonon relaxation}

The e-ph relaxation length is found in a standard way from the cooling rate as $l_{e-ph} = \sqrt{ \sigma \mathcal{L}/(n \Sigma_{e-ph}T_e^{n-2}) }$, where $n=5$ is the assumed exponent in the heat balance equation and $\Sigma_{e-ph}=4.8\,\text{nW}\mu\text{m}^{-3}\text{K}^{-5}$ is taken from measurements of the critical Joule power in Type-II devices in $B = 0$. $\mathcal{L}=(\pi^2/3)(k_B/e)^2=2.44\cdot 10^{-8}$ V$^2$K$^{-2}$ is the Lorenz number. In this way we obtain $l_{e-ph} = 20 ~\mu$m at $T_e=0.45$\,K, which corresponds to $\tau_{e-ph}=l_{e-ph}^2/D=62$ ns.


\begin{figure}[t!]
	\begin{center}
		\vspace{0mm}
		\includegraphics[width=1\linewidth]{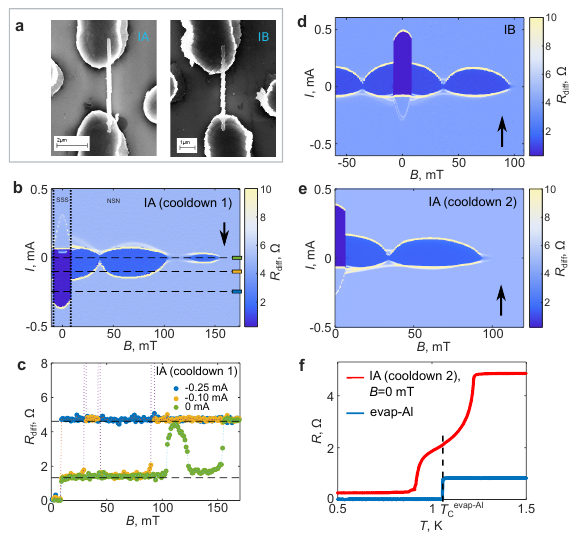}
	\end{center}
	\caption{ (a) Scanning electron micrographs of the devices IA and IB. (b),(e) The differential resistance versus current and $B$-field for the IA device in different cooldowns. (d) the same for the device IB. Black arrows indicate the direction of the current sweep.
		(c) Representative $R_\mathrm{diff}$-$B$ curves measured at the fixed current, corresponding to the cross-cuts in panel (b). The wiring resistance of $R_\mathrm{w}=0.26\,\mathrm{\Omega}$ is subtracted from all data. The horizontal dashed lines correspond to  $R_\mathrm{int}=1.33\,\mathrm{\Omega}$ and $R_\mathrm{int}+R_\mathrm{sh}=4.62\,\mathrm{\Omega}$. 
		(f) $R(T)$ dependencies in $B=0$  for the device IA in cooldown 2 (red line) and for the test structure (blue line).
	 } 
	\label{FigS2}
\end{figure}

\section{Additional Experimental data}

\subsection{Type-I devices}
Here we supplement the description of the type-I devices by adding device images and differential resistance data in Fig.~\ref{FigS2}. Scanning electron micrographs of the devices IA and IB are shown in Fig.~\ref{FigS2}a.  Color-scale plots of the differential resistance are shown in  Figs.~\ref{FigS2}b, ~\ref{FigS2}d and ~\ref{FigS2}e for different samples, cooldowns and/or current sweep directions (see legends). 
Representative dependencies $R_\mathrm{diff}(B)$ corresponding to the data of Fig.~\ref{FigS2}b are shown in Fig.~\ref{FigS2}c.

We also performed a cross-check experiment to test the resistivity, critical temperature and critical current of the contact pads in type-I devices. Here, we fabricated four-terminal devices consisting of 100~nm thick Au layer and 350~nm thick Al layer evaporated on top. 
The test-structure had a rectangular shape of the length of $300\,\mu$m and the width of $2\,\mu$m. The measured  $T$-dependence of the linear response resistance is shown in Fig.~\ref{FigS2}f by the blue line. The measured resistance per square equals $R_{\square} \approx 5.5\times10^{-3}\,\Omega$. This allows to estimate the normal state resistance of two contact pads in type-I devices at $\approx0.15\,\Omega$, which is negligible compared to the interface contribution $R_\mathrm{int}=1.33\,\mathrm{\Omega}$. The measured critical temperature of the test structure is $T_\mathrm{c}\approx1.05$~K. This $T_\mathrm{c}$ is slightly higher than that of the evap-Al in type-I devices ($\approx 0.95$\,K), see the low-$T$ step on the $R(T)$ curve in the device IA (red line in Fig.~\ref{FigS2}f). This discrepancy reflects the fact that the Au layer was thicker in type-I devices ($\approx150$\,nm), which resulted in stronger inverse proximity effect. The critical current of the test structure exceeded 20~mA and was even difficult to measure because of the self-heating of the wiring.        

\subsection{Type-II}
\begin{figure}[t!]
	\begin{center}
		\vspace{0mm}
		\includegraphics[width=1\linewidth]{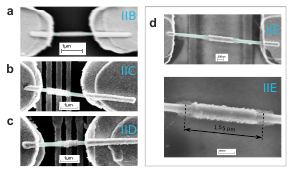}
	\end{center}
	\caption{Scanning electron micrographs of devices IIB-IIE with false colored InAs parts. } 
	\label{FigS3}
\end{figure}

\begin{figure}[t!]
	\begin{center}
		\vspace{0mm}
		\includegraphics[width=1\linewidth]{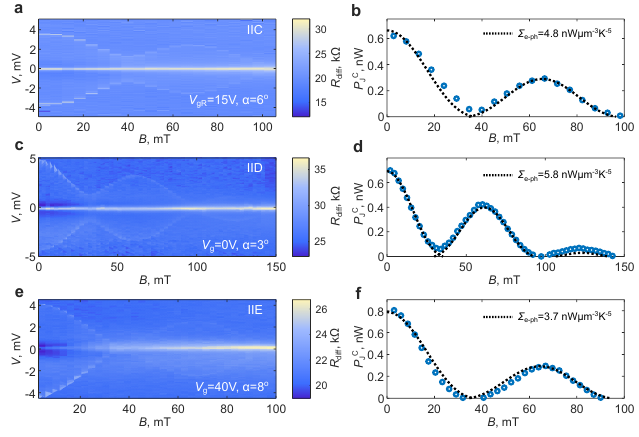}
	\end{center}
	\caption{(a) the color-scale plot of $R_\mathrm{diff} (B,V)$ for the device IIC. The gate voltage on the right most local back gate $V_\mathrm{gR}$ and the angle $\alpha$ between the NW axis and the magnetic field are indicated in the legend.
		(b) The $B$-dependence of the critical Joule power for the same device (symbols) and the model fit to the heat balance equation (dotted line). (c)-(d) the same for the device IID. The global back gate voltage $V_\mathrm{g}$ and the angle $\alpha$ are indicated in the legend.
		(e)-(f) the same for device IIE. The global back gate voltage $V_\mathrm{g}$ and the angle  $\alpha$ are indicated in the legend.} 
	\label{FigS4}
\end{figure}

Fig.~\ref{FigS3} present the scanning electron micrographs of the devices IIB-IIE with partially etched shell. The device IIB is equipped with the global back gate, see Fig.~\ref{FigS3}a, whereas other devices have both the global back gate and several local back gates, represented by thin metal strips underneath the NW, see Figs.~\ref{FigS3}b-~\ref{FigS3}d. In the device IIC, the right most local back gate was used to tune the resistance (Figs.~\ref{FigS3}b), while other gates were grounded. In all other devices the local gates were grounded and the global back gate was used.

Figs.~\ref{FigS4}a,~\ref{FigS4}c and~\ref{FigS4}e  present color-scale plots of $R_\mathrm{diff}$ as a function of the $B$-field and bias voltage, respectively, in devices IIC, IID and IIE. The critical Joule power $P_\mathrm{J}^\mathrm{C}(B)$ is extracted from this data and plotted in Figs.~\ref{FigS4}b,~\ref{FigS4}d and~\ref{FigS4}f (circles). The dotted lines represent the fits of $P_\mathrm{J}^\mathrm{C}(B)$ to the heat balance model, described in the main text.
 
The fit procedure is performed as follows. We determine the volume of the epi-Al $\mathcal{V}_\mathrm{Al}$ from the device images and obtain the electron-phonon (e-ph) cooling rate $\Sigma_\mathrm{e-ph}$ from the value of the critical Joule power in $B=0$: $ \Sigma_\mathrm{e-ph} =P_{J}^{C}(B=0)/2\mathcal{V}_\mathrm{Al}\left[(T_\mathrm{c}^0)^5 - T_0^5\right]$. This allows to get rid of the uncertainty related to the unknown angle $\alpha$ between the NW axis and the magnetic field. Next, the same heat balance equation and the obtained value of  $\Sigma_\mathrm{e-ph}$ are used to obtain the $B$-dependence of the superconducting critical temperature of the epi-Al:
\begin{align}	
	T_\mathrm{c}(B) = \sqrt[5]{ T_0^5 +\frac{P_\mathrm{J}^\mathrm{C}(B)}{2\Sigma_\mathrm{e-ph}\mathcal{V}_\mathrm{Al}}} \label{Joule_fit}
\end{align}	

The dependence $T_\mathrm{c}(B)$ is fitted with the Abrikosov-Gorkov equation (\ref{AG_eq}) using $\alpha$ and $\rho_\mathrm{i}$ as the fit parameters. The final step is to calculate $P_\mathrm{J}^\mathrm{C}(B)$ from the inverted expression (\ref{Joule_fit}). The fit parameters for all our type-II devices are summarized in Table~\ref{table}.

\begin{table}[h!]
	\centering 
	\caption{ Parameters of the type-II devices. }          
	\begin{tabular}{|c|c|c|c|c|c|c|}\hline Device &  $\rho_\mathrm{i}$, nm & $L$, $\mu$m & $\mathcal{V}_\mathrm{Al}$, $\mu$m$^3$ & $\alpha$,$^o$ & $\Sigma_\mathrm{e-ph}$, $\frac{nW}{ \mu m^{3} K^{5} }$ \\\hline
	IIA & 76 & 1.0	& 0.026 & 6 & 4.8 \\
	IIB & 84 & 1.85	& 0.051 & 5 & 4.3 \\
	IIC & 76 & 0.97	& 0.025 & 6 & 4.8 \\
	IID & 80 & 0.82	& 0.022 & 3 & 5.8 \\
	IIE	& 76 & 1.55 & 0.040 & 8 & 3.7 \\\hline
\end{tabular} \label{table}
\end{table}

\subsection{Shot noise measurements}
\begin{figure}[t!]
	\begin{center}
		\vspace{0mm}
		\includegraphics[width=1\linewidth]{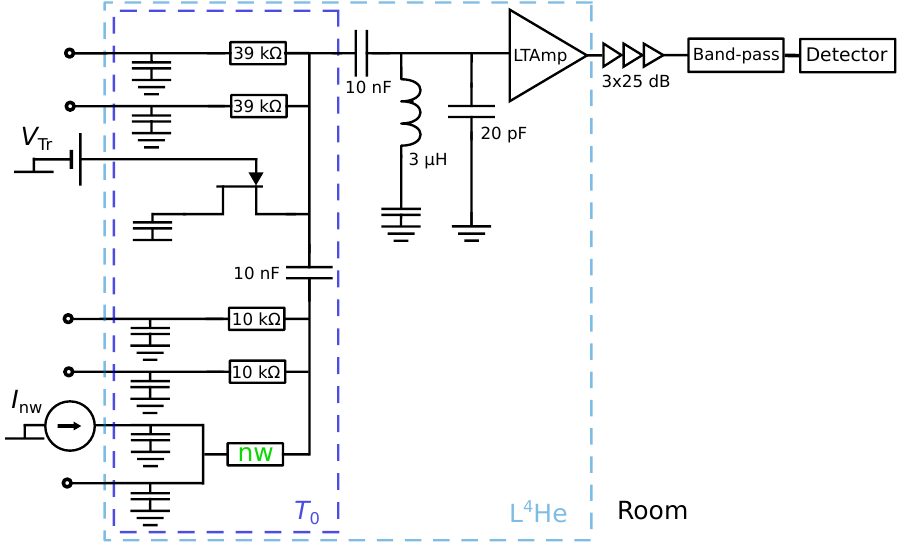}
	\end{center}
	\caption{ The transport and shot noise measurement scheme. } 
	\label{FigS5}
\end{figure}

Shot noise measurements in type-II devices were carried out in a setup schematized in Fig.~\ref{FigS5}. The NW device is embedded into a resonant tank circuit at the input of a home-made low-temperature amplifier (LTAmp). The resonance frequency is determined by the inductance ($\approx3\,\mu$H) and cable capacitance ($\sim20\,$pF) at the input of the LTAmp and is about 20~MHz, slightly varying in different setups used. Apart from the NW device, the scheme includes four load resistors (RF grounded on one side by means of 10~nF capacitors). These resistors serve for DC measurements. The key part of the setup is the commercial transistor ATF-55143, which is used to calibrate the gain of the noise measurement scheme via Johnson-Nyquist thermometry. This procedure also allows to determine the input current noise of the LTAmp, which has a typical value of $\sim2\times10^{-27}\,\mathrm{A}^2/\mathrm{Hz}$. The calibration is performed at two bath temperatures of 4.2\,K and $T_0\approx0.5$\,K. The details of the calibration and data analysis are similar to those described in supplemental materials of Refs.~\cite{tikhonov2014a,denisov2022}. 

Here we explain the fitting procedure of the shot noise data in type-II devices, presented in Fig.~3f in the main text. In this experiment, the central epi-Al island is in the normal state. Strong e-ph relaxation is assumed, which results in the local equilibrium EED with the electronic temperature of $T_\mathrm{e}$. The electronic temperature is determined from the heat balance equation of the kind (\ref{Joule_fit}) with $T_\mathrm{e}$ substituting the $T_\mathrm{c}(B)$ and Joule power substituting $P_\mathrm{J}^\mathrm{C}(B)$. In this way, the dependence $T_\mathrm{e}(V)$ is obtained in each device. The spectral density ($S_\mathrm{I}$) of the current noise of the NW device is obtained from the spectral densities of current noises of the left ($S_\mathrm{I}^\mathrm{L}$) and right  ($S_\mathrm{I}^\mathrm{R}$) InAs segments in a usual way~\cite{dejong1996}:
\begin{align}	
	S_\mathrm{I} = \frac{ S_\mathrm{I}^\mathrm{L} R_\mathrm{L}^2 + S_\mathrm{I}^\mathrm{R} R_\mathrm{R}^2 }{R^2}=\frac{a^2S_\mathrm{I}^\mathrm{L} + S_\mathrm{I}^\mathrm{R}}{(a+1)^2},	
\end{align}	
where resistances $R_\mathrm{L}$, $R_\mathrm{R}$ correspond to the to left and right InAs segments of the NW device, $R=R_\mathrm{L}+R_\mathrm{R}$ is the total device resistance and $a=R_\mathrm{L}/R_\mathrm{R}$. As discussed in the main text, the device resistance changes at finite bias, as a result of Coulomb interaction effects~\cite{nazarov1999,golubev2001,yeyati2001}. For simplicity, we assume that the resistance ratio $a$ is bias-independent. The spectral density of the current noise of the left  and right InAs segments reads:
\begin{align*}	
	S_\mathrm{I}^\mathrm{L,R} = \frac{2k_\mathrm{B}(T_\mathrm{e}+T_0)}{R_\mathrm{L,R}}  + \frac{F}{R_\mathrm{L,R}} \int dE \left[f_\mathrm{Al}(E) - f_\mathrm{L,R}(E)\right]^2,
\end{align*}	
where $f_\mathrm{Al}(E)=f_0(E,T_\mathrm{e})$ is the EED in the epi-Al island, $f_\mathrm{L}(E)=f_0(E+aV/(a+1),T_0)$ is the EED in the left terminal and $f_\mathrm{R}(E)=f_0(E-V/(a+1),T_0)$ is the EED in  the right terminal. In this equation the change of NW resistance with bias voltage is taken into account as $R_\mathrm{L}=aV/(a+1)I$ and $R_\mathrm{R}=V/(a+1)I$, where $V$
 and $I$ are, respectively, the measured bias voltage and the current in the NW device. For simplicity, we assume that the Fano factors $F$ of both InAs segments are the same, but allowed a slight variation of $F$ among different samples. The fits in Fig.\,3f in the main text are obtained with $F=1/3$ and $a=0.485$ in device IIA, $F=0.25$ and $a=0.36$ in device IIB and $F=1/3$ and $a=0.4$ in device IIC. The choice of $F\approx1/3$ seems reasonable, since our devices are usually close to the diffusive transport regime. The choice of $a$ is justified by the quality of the shot noise fits, however, it is difficult to verify independently in our two-terminal configuration.


%